\documentclass[a4paper,fleqn,usenatbib]{aa}  

\usepackage{graphicx}
\usepackage{txfonts}
 \usepackage[colorlinks,citecolor=blue,linkcolor=blue,anchorcolor=blue]{hyperref}
\usepackage{amsmath}
\usepackage{amssymb}
\usepackage{breqn}
\usepackage{gensymb}

\newcommand{\rg}{GM/c^2}
\newcommand{\tunit}{GM/c^3}

\newcommand{\msun}{\mathrm{M}_{\odot}}

\newcommand{\mdotu}{\rm M_\odot yr^{-1}}

\newcommand{\trat}{T_{\rm p}/T_{\rm e}}

\begin{document} 
\title{General relativistic magnetohydrodynamical 
$\kappa$-jet models for Sgr A*}
\titlerunning{GRMHD $\kappa$-jet models for Sgr A*}
\authorrunning{J. Davelaar et al.}

\author{J. Davelaar\inst{1},
M. Mo{\'s}cibrodzka\inst{1},
T. Bronzwaer\inst{1}, and
H. Falcke\inst{1}\inst{2}}

\institute{
Department of Astrophysics/IMAPP, Radboud University Nijmegen, P.O. Box
9010, 6500 GL Nijmegen, The Netherlands \\email: {\tt j.davelaar@astro.ru.nl}
\and
ASTRON, Dwingeloo, The Netherlands
}

   \date{Received XX, YY; accepted XX, YY}

 
  \abstract
    {
The observed spectral energy distribution of an accreting supermassive black hole
typically forms a power-law spectrum in the Near Infrared (NIR) and optical wavelengths, that may be interpreted as a
signature of accelerated electrons along the jet. However, the details of
acceleration remain uncertain.}
{In this paper, we study the radiative properties of jets produced in axisymmetric GRMHD simulations of
hot accretion flows onto underluminous supermassive black holes both
numerically and semi-analytically, with the aim of investigating the differences between models with and without accelerated electrons inside the jet. }
{We assume that electrons are accelerated
in the jet regions of our GRMHD simulation. To model them, we modify the electrons' distribution function in the jet regions from a purely relativistic thermal distribution to a combination of a relativistic thermal distribution and the
$\kappa$-distribution function (the $\kappa$-distribution function is itself a combination of a relativistic thermal and 
a non-thermal power-law distribution, and thus it describes accelerated~ electrons). Inside the disk, we assume a thermal distribution for the electrons. In order to resolve the particle acceleration regions in the
GRMHD simulations, we use a coordinate grid that is optimized for modeling jets. We
calculate jet spectra and synchrotron maps by using the ray tracing code {\tt
RAPTOR}, and compare the synthetic observations to observations of Sgr~A*. 
Finally, we compare numerical models of jets to semi-analytical ones.}
{We find that in the $\kappa$-jet models, the radio-emitting region size, radio flux, and 
spectral index in NIR/optical bands increase for decreasing values of the $\kappa$
parameter, which corresponds to a larger amount of accelerated electrons. This is in agreement with analytical predictions. 
In our models, the size of the emission region depends roughly linearly on the 
observed wavelength $\lambda$, independently of the assumed distribution function. 
The model with $\kappa = 3.5$, $\eta_{\rm acc}=5-10\%$ (the percentage of electrons that are accelerated), and observing angle $i = 30^\degree$ fits the
observed Sgr~A* emission in the flaring state from the radio to the NIR/optical regimes, while $\kappa = 3.5$, $\eta_{\rm acc}< 1\%$, and observing angle $i = 30^\degree$ fit the upper limits in quiescence. At this point, our models (including the purely thermal ones) cannot 
reproduce the observed source sizes accurately, which is probably due to the assumption of 
axisymmetry in our GRMHD simulations. The $\kappa$-jet models naturally recover the observed nearly-flat radio 
spectrum of Sgr~A* without invoking the somewhat artificial isothermal jet model that 
was suggested earlier.  
 }
 {From our model fits we conclude that between $5\%$ to $10\%$ of the electrons inside the jet of Sgr~A* are accelerated into a $\kappa$ distribution function when Sgr~A* is flaring. In quiescence, we match the NIR upper limits when this percentage is less than $1\%$.}

   \maketitle
%

\section{\label{sec:level1} Introduction}

In general relativistic magnetohydrodynamics (GRMHD) global simulations
of weakly radiating accretion flows onto a black hole, the electron energy
distribution function is not explicitly modeled.  In these simulations, the
accreting plasma is collisionless (i.e., the Coulomb mean free path for
electrons is much larger than $\rg$), which means that the electrons are decoupled
from the dynamically important, more massive protons. The processes that control the
electron distribution function, such as magnetic reconnection, dissipation of
turbulent energy, shocks, and/or other plasma effects that particle-in-cell
simulations show, cannot be resolved with the current computational grids used
in global simulations of the accretion flows. To predict the radiative
properties of GRMHD accretion flows, and to improve the predictive power of the theory of accretion
with respect to observations, sub-grid models for electron heating and acceleration have to be invoked.

Sagittarius~A* (Sgr~A*) is a supermassive black hole system that allows one to
observationally test the aforementioned GRMHD models of accretion flows
(\citealt{bhc}). Millimeter-Very Long Baseline Interferometry (mm-VLBI) is capable of resolving the shadow of the event horizon (\citealt{falckemelia}), making this an ideal laboratory not only to tests Einstein's General Theory of Relativity (GR) but also to investigate electron acceleration in the vicinity of a black hole. Most of the radiative models for Sgr~A*, which are based on post-processing GRMHD simulations,
assume that electrons have a thermal, relativistic (Maxwell-J{\"u}ttner)
distribution function, and that the proton-to-electron temperature ratio is
constant across the simulation domain (\citealt{goldstone2005},
\citealt{noble2007}, \citealt{sgra2009}, \citealt{jason_sgra},
\citealt{dexter2012}, \citealt{Shcherbakov2012}). When the proton-to-electron temperature is constant, the disk dominates the images and spectra since most of the matter resides there. We have recently extended
these radiative models by making the temperature ratios a function of the plasma
$\beta$ parameter, where $\beta = \frac{P_{\rm gas}}{P_{\rm B}}$ is the ratio
of gas to magnetic pressures. In these extended models, the electrons are
hotter in the more magnetized plasma, which is usually outflowing from the
system. The reason for this is that the previously mentioned models do not recover the flat radio spectra. The $\beta$ parameterization enforces that the disk emission is suppressed by significantly decreasing the temperature of the electrons in those regions. As a consequence of this, the jet will be the dominant source of emission. These modifications to the electron temperature model allowed us to recover some
basic observational characteristics of Sgr~A* (a roughly flat radio spectral slope
and a size vs. wavelength relationship that is in agreement with observations) (\citealt{sgra},
\citealt{sgra-3d}, \citealt{chan_sgra}, \citealt{chang_sgra2}, \citealt{gold2017}). 
Our model for the electron temperatures as a function of the $\beta$ plasma parameter 
is now roughly confirmed with extended-GRMHD simulations that self-consistently 
take into account the evolution of the electron
temperatures (\citealt{ressler}, \citealt{ressler2017}). Moreover, GRMHD
simulations with the new electron temperatures can naturally explain the
symbiosis of disks and jets observed in many accreting black hole systems
(\citealt{falcke}, \citealt{Monika2016b}).

Observations of Sgr~A* show flares in the NIR/optical wavelengths with a spectral index of $\alpha \approx -0.7 \pm 0.3$ (\citealt{bremer2011}). These flares are indicators of accelerated non-thermal electrons in the accretion flow, which is not accounted for in our previous models of Sgr~A*.

Due to computational constraints, it is challenging to make a first-principles
model for particle acceleration in GRMHD simulations (but see \citealt{chael2017}). A simpler approach can
be adopted in which the non-thermal particles 
are included in a phenomenological prescription.
The accelerated electrons can be described by a hybrid
distribution function that is constructed by 'stitching' a power-law tail onto a thermal distribution function. The hybrid distribution function is then described by a few free parameters: the power law index ($p$), the
acceleration efficiency ($\eta$, which is the amount of energy in the accelerated electrons compared to the total energy budget), and the maximum
Lorentz factor ($\gamma_{max}$) or the Lorentz factor at which radiative cooling starts
to dominate ($\gamma_{c})$. The minimum Lorentz factor ($\gamma_{min}$) is
then calculated at the `stitching' point.  With an underlying model for the accreting plasma, these free parameters can be then constrained by comparing the model emission to the observational data.

One of the first attempts to model electrons around Sgr~A*
with the hybrid distribution function is presented in \cite{ozel}. Their underlying model for the accreting plasma is a semi-analytical radiatively inefficient accretion flow (RIAF) (\citealt{narayan1994}, \citealt{narayan1995a}, \citealt{narayan1995b}, and \citealt{chen1995}). \cite{ozel} found that the observed low-frequency shoulder of the Sgr~A* spectrum is well described by emission from RIAF electrons described by a hybrid distribution function with $\eta\approx 0.01$ and $p=3-3.5$. Similar conclusions were later reached by, e.g., \citealt{yuan2003} and \citealt{broderick2016}.

Recently, \cite{mao} studied the effects of accelerated electrons in GRMHD simulations
on the mock spectra and millimeter images of Sgr~A* using either a hybrid distribution function
(with $p=3.5$ and various values of $\eta$) or a multi-Maxwellian distribution function. They found that the accelerated, 
high energy electrons not only alter the observed spectral energy distribution
shape but also lead to more extended and diffuse resolved millimeter
images of the source in the case of the hybrid distribution function. This has a few interesting implications for interpreting
the VLBI observations of Sgr~A* (\citealt{bower2004}, \citealt{shen2005}, 
\citealt{doeleman}, \citealt{bower2014}, \citealt{brinkerink2016}).
Similarly to early semi-analytical model by \citet{ozel},
\citet{mao} assumed constant acceleration parameters in the entire simulation
domain, which is reasonable but does not have to be the case. For example,
\citet{ball} insert accelerated electrons in low-$\beta$ regions where particle acceleration is expected to occur via magnetic
reconnection) of GRMHD
simulations of Sgr~A*, and study the impact of accelerated electrons on the emitted SED with
the goal to explain near-infrared (NIR) and X-ray flares observed in Sgr~A*. Their best fit model assumes the electron acceleration efficiency
$\eta=0.1$ and a power-law index $p=3.5$. A similar approach 
(with $p=3-3.5$ and acceleration efficiency proportional to magnetic energy) 
was earlier adopted by \citet{dexterM87} to model the size of the near-horizon
emission in M87 radio core (hereafter M87*).

In this paper, we study the effects of particle acceleration on spectra and images of axisymmetric GRMHD models of accretion flows with jets. The goal is to extend our current models with electron acceleration to see if it is possible to obtain the nearly flat SED of Sgr A* and set constraints on the amount of electron acceleration during NIR/optical flares. Our underlying accretion model assumes that the proton-to-electron temperature ratio is a function of the plasma $\beta$ parameter and that electrons are hotter in low-$\beta$ regions of the simulations that are associated with the jet outflow. We model emission from radio to NIR/optical frequencies.
The main source of photons in the magnetized, relativistically hot plasma studied here is the synchrotron process. We ignore inverse-Compton scatterings, hence do not model the  X-ray emission.
 
The accelerated electrons investigated in this paper are described by the $\kappa$
distribution function \citep{Vasyliunas} instead of the hybrid distribution
function. The $\kappa$ distribution function smoothly connects the thermal core to the
power-law tail (which is not the case in the hybrid model), and better
describes processes such as first-order Fermi acceleration
\citep{Livadiotis2013}. The derivation of the function
can be found in \citet{kappa}. The $\kappa$ distribution function is related to the thermal distribution function as a limit
\begin{equation}
\begin{aligned}
f_{\rm thermal}(\Theta_{\rm e},\gamma) & \propto e^{-(\gamma-1)/\Theta^2}\\ &
=\lim_{\kappa
  \rightarrow \infty} \left(1+\frac{\gamma - 1}{\kappa
  \Theta_{\rm e}}\right)^{-\kappa -1} \propto \lim_{\kappa
  \rightarrow \infty} f_\kappa (\Theta_{\rm e},\gamma),
  \end{aligned}
\end{equation}
where $\kappa$ is a free parameter of the distribution function, and $\Theta_{\rm e}$ is the dimensionless temperature of the electrons involved. In the power-law part of the distribution function, the
parameter $\kappa$ is related to the power-law index $p$ by
$\kappa=p+1$, such that in the limit of $\gamma \gg 1$ the $\kappa$ distribution
function asymptotically approaches $f_\kappa (\Theta,\gamma) \propto \gamma^{-p}$. The $\kappa$ distribution function has been used to describe plasma in the solar
wind \citep{decker} and plasmas in, e.g., coronal flares on the Sun
\citep{Livadiotis2013}. Theoretically \cite{kunz2016} found that the
distribution function of accelerated particles in accreting systems follows a
$\kappa$ distribution.

The paper is organized as follows. In Sections~\ref{sims} and~\ref{numgrid},
we explain how the GRMHD simulations are set up. In section~\ref{kappa}, we
describe radiative transfer parameters and the acceleration of electrons. We present and discuss the results in Sections~\ref{results}
and~\ref{discuss}, respectively.

\section{\label{sec:level2}Methods}

\subsection{GRMHD simulation}\label{sims}

Our accreting plasma model is based on GRMHD simulations of magnetized gas
around a supermassive, spinning black hole. The simulation
begins with a torus in hydrostatic equilibrium in a Keplerian orbit around
a Kerr black hole \citep{FMtorus}. The size of the initial torus is set by two
parameters: the inner edge of the torus $r_{\rm in}=6 \,\rg$, and the radius $r_{\rm max}=12 \,\rg$ of the pressure maximum of the torus, where $\rg$ is the simulation length unit. We evolve the flow with the GRMHD code {\tt HARM2D}, where we used the standard setup for reconstruction schemes and constrained transport as described in \cite{HARM}.

The initial torus is seeded with a weak poloidal magnetic field where the topology
follows the isodensity contours of the torus. The strength of the magnetic field
is set via the dimensionless plasma $\beta$ parameter defined as:
\begin{equation}
\beta = \frac{P_{gas}}{P_{mag}}=\frac{ u (\gamma_{\rm ad} - 1)}{B^2},
\end{equation}
where $\gamma_{\rm ad}$ is the adiabatic index, $u$ is the internal energy density, and
$B$ is the magnetic field strength. The initial torus has a minimum $\beta =100$;
in other words, initially, the magnetic fields are relatively weak.

t

We are interested in modeling $\nu= 10^9-10^{15}$ Hz
emission originating from the inner accretion flow (high-energy end of the spectrum) and the extended jet (low-energy end of the spectrum).
Models of low-frequency emission from the jet require the simulation to be radially extended to an outer radius of $r_{\rm out}=1000 \,\rg$. We evolve the
simulation until the final time $t_{\rm f}=4000 \,\tunit$, where $\tunit$ is the simulation time unit. This $t_{\rm f}$ allows the  jet to reach the outer boundary of the computational domain. The simulation duration corresponds to
15 orbital periods of the torus.

\subsection{Numerical grid for simulating disks and jets}\label{numgrid}

\begin{figure}
\centering \includegraphics[width=0.5\textwidth]{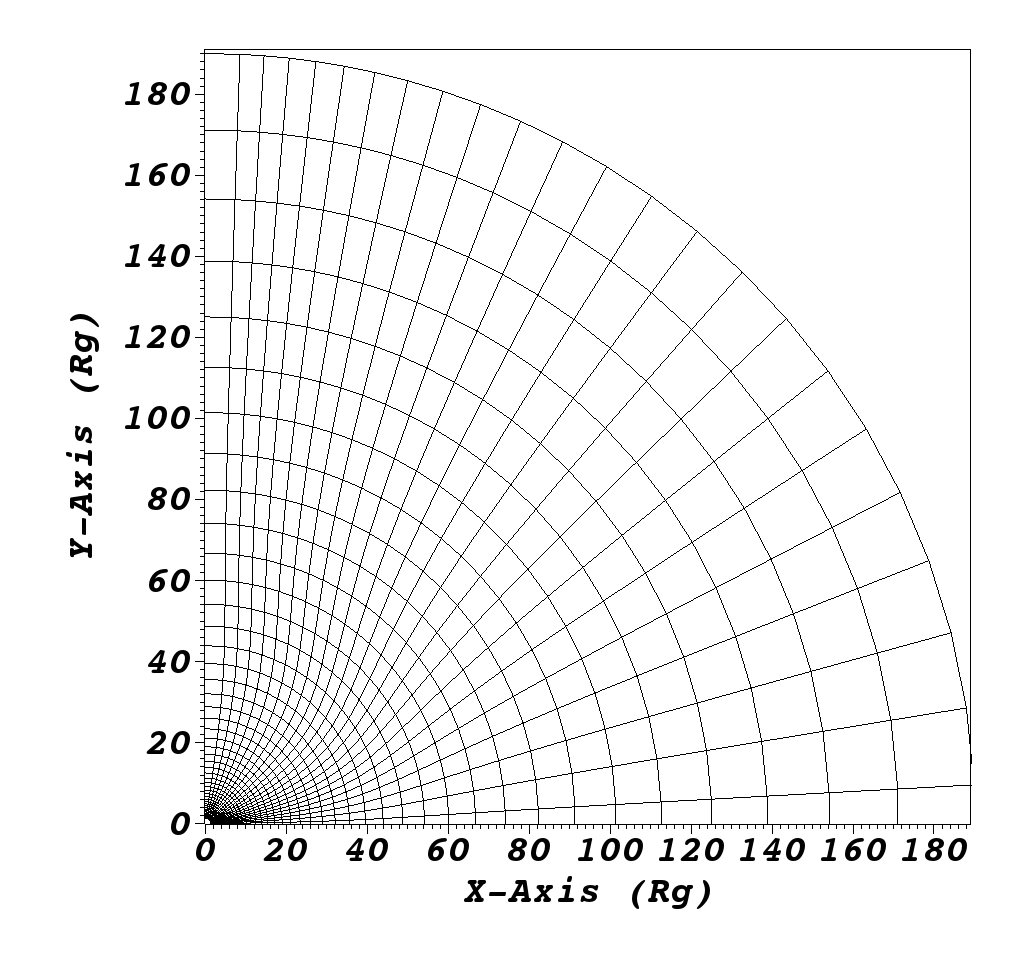}
\caption{The upper half of an MMKS coordinate system that focuses the grid
  resolution in the polar regions. For clarity, a lower resolution grid is
  displayed.}\label{grid}
\end{figure} 

The dynamical simulation is carried out in mixed modified Kerr-Schild (MMKS)
coordinates (\citealt{noble2007}) ($X_0,X_1,X_2,X_3$), which are related to
standard Kerr-Schild coordinates ($t,r,\theta,\phi$) by:

\begin{align}
t&=X_0,\\ r&= e^{X_1}, \\ \theta&= \pi X_2 + \frac{2 h_{\rm slope}}{\pi}\sin\left(2 \pi
X_2\right)\arctan\left(s\left(x_0-X_1\right)\right), \\ \phi&=X_3,
\end{align}
where $h_{\rm slope}$, $s$, and $x_{\rm 0}$ are free parameters of the coordinates system that can be used to refine the coordinate grid near the equatorial plane, where the most dense region of the disk resides, and in the polar regions where a jet is expected to form. The parameter $h_{\rm slope}$
controls the grid spacing near the equatorial plane in the innermost region
of the simulation. The parameter $x_0$ is defined as $ x_0 \equiv log(r_{\rm tr})$, where $r_{\rm tr}$ 
is a transition radius at which the grid transitions from a parabolic to a conical shape along the jet axis, and $s$ defines how rapidly this transition occurs.

The simulation is performed in two dimensions, with a grid resolution of
$N_{\rm X_1} \times N_{\rm X_2}=512 \times 528$ and grid parameters $h_{\rm slope}=0.35$,
$s=2$, and $r_{\rm tr}=50\,\rg$. A visualization of the MMKS coordinate system
is presented in Figure~\ref{grid}.

\subsection{Radiative transfer model and electron distribution
  functions}\label{kappa}

The SEDs and images of the GRMHD accretion flow models are computed using general-relativistic, ray-tracing radiative transport scheme {\tt RAPTOR} \citep{raptor}.

The GRMHD simulations are scale-free, this means that the quantities obtained are unitless. Calculation of mock observations of these models requires scaling them to c.g.s. units. The scaling depends on observational constraints such as distance, the mass of the black hole, and the matter content of the accretion disk.  The simulation length unit is ${\mathcal L}=\rg$, the time unit is ${\mathcal T} = \tunit$, and the mass unit is ${\mathcal M}$. While a good estimate of the black hole mass, $M$, exists for Sgr~A* (hence the
length unit [cm] and time unit [s] are reasonably well-known), the accretion mass unit,
${\mathcal M}$ is a free parameter of the system, which has to be constrained
by fitting our model spectrum to observations. The parameter $\mathcal{M}$
determines the density of the accretion flow, and thus the mass accretion
rate onto the black hole. The dimensionless accretion rate $\dot{M}_{\rm sim}$
can be converted to the accretion rate in c.g.s. units by
$\dot{M} = \dot{M}_{\rm sim} {\mathcal M} {\mathcal T}^{-1}$. To convert the plasma density, specific internal energy, and magnetic field strength from code units to c.g.s. units we use the following scaling factors: $\rho_0 =
\mathcal{M}/\mathcal{L}^3$, $u_{0}=\rho_0 c^2$, and $B_0=c\sqrt{4\pi \rho_0}$.

In the GRMHD simulation, we only evolve protons. We, therefore, need assumptions for the electron distribution functions and how the density and temperature of the electrons depend on the computed plasma variables.

We divide the simulation volume into three regions: the disk; the jet-sheath; and the jet-spine.
In each region, we assume different electron distribution functions.  In the
disk region, we assume that electrons have a thermal (Maxwell-J{\"u}ttner)
distribution function. We accelerate electrons in the jet-sheath region,
defined using the Bernoulli parameter $Be = -h u_{\rm t} > 1.02 $ \citep{sgra} with
$h$ the gas enthalpy and $u_{\rm t}$ the time component of the four-velocity. We
neglect any emission from the the jet-spine region, defined using the
magnetization parameter $\sigma = \frac{B^2}{\rho}$. The matter content and
energy content of jet-spine is set by numerical floor values. These numerical
floor values can result in unphysical large fluctuations in temperature that must be excluded from
the synthetic images. This is caused by the conservative nature of the scheme that HARM2D uses; the magnetic energy is large while the internal energy is low, therefore, tiny fluctuations of the magnetic energy can result in large fluctuations of the internal energy because the codes will enforce conservation of energy. This behavior in the jet spine can be found in the regions close to $\theta =0$ and $\pi$ where  $\sigma>1.0$, ~any radiation from these regions is ignored.

\begin{figure}
\centering \includegraphics[width=0.45\textwidth]{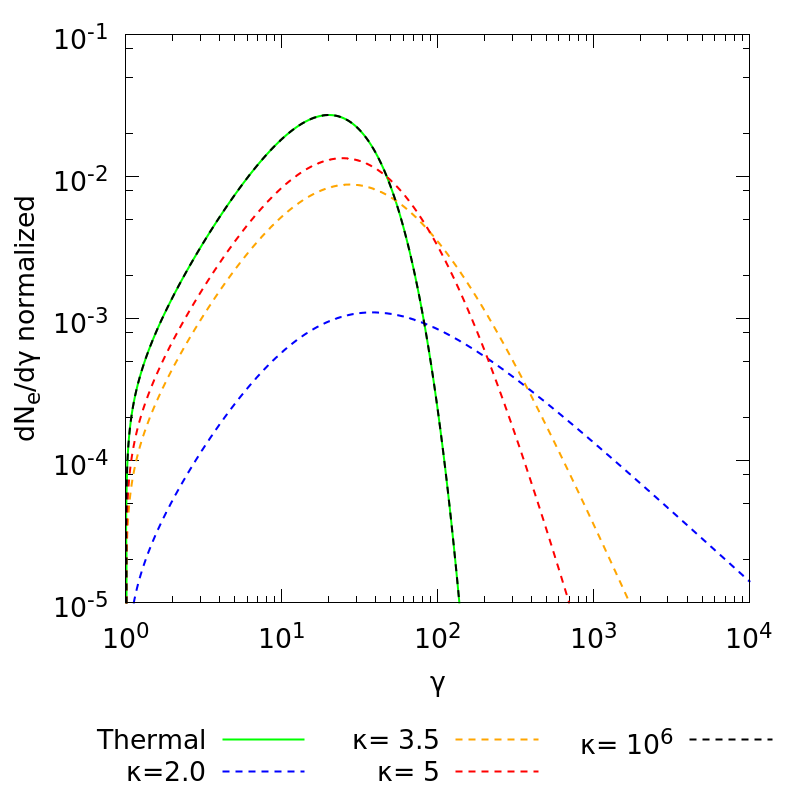}
\caption{The $\kappa$ distribution function for different values of the
  $\kappa$ parameter (dashed lines). In the limit of large $\kappa$ (black
  dashed line) the relativistic thermal distribution function is recovered
  (green solid line).\label{distr}}
\end{figure} 

The electron temperature,
$T_{\rm e}$, is computed assuming that the proton-to-electron coupling
depends on plasma magnetization (\citealt{M87};
\citealt{monika2017}). We use the following law for the coupling of the electron
and proton temperatures:
 \begin{equation}\label{Eq:Te}
 \frac{T_{\rm p}}{T_{\rm e}} = R_{\rm low}\frac{1}{1+\beta^2} + R_{\rm high}
 \frac{\beta^2}{1+\beta^2},
 \end{equation}
where $\beta= \frac{P_{\rm gas}}{P_{\rm mag}}$ is the ratio of the gas
pressure to the magnetic field pressure $P_{mag}=B^2/2$. $R_{low}$ and $R_{high}$
are free parameters. In a strongly magnetized plasma, $\beta \ll 1$ and
$\trat \rightarrow R_{\rm low}$. In a weakly magnetized plasma, $\beta \gg 1$
and so $\trat \rightarrow R_{\rm high}$. We set $R_{\rm low}=1$ and $R_{\rm
  high} = 25$ so that the electrons are cooler in the disk and hotter
toward the jet.

The energy distribution function of accelerated electrons in the jet-sheath is
described by a combination of a thermal distribution and the relativistic $\kappa$ distribution function
(\citealt{kappa}):

\begin{equation}
\frac{dn_{\rm e}}{d\gamma} = N \gamma \sqrt{\gamma^2 -1} \left( 1 +
\frac{\gamma -1}{\kappa \Theta}\right)^{-(\kappa + 1)},
\end{equation}
where $\kappa$ is a free parameter, $\Theta$ is the dimensionless
temperature defined as $\Theta_{\rm e} \equiv k_bT_{\rm e}/m_{\rm e}c^2$, and $N$ is a normalization ( that depends on $\Theta_{\rm e}$ and $\kappa$) such that 
$\int_1^\infty \frac{dn_{\rm e}}{d\gamma} d\gamma = n_{\rm e}$. 
The $\kappa$ distribution
function consists of a non-thermal power-law tail that smoothly connects to a
thermal-like core. In the limit of $\kappa \rightarrow \infty$, the $\kappa$
function asymptotically approaches the thermal distribution function with temperature
$\Theta_{\rm e}$. Figure~\ref{distr} shows
the $\kappa$ distribution function for a few values of the $\kappa$ parameter.

For large $\gamma$, the distribution function asymptotically approaches a
power-law with index $p$ that is related to $\kappa$ by $\kappa \equiv p +
1$. Hence, the spectral index $\alpha$ of the optically thin part of the
observed spectrum (where the observed flux density is $F_{\nu}\propto
\nu^\alpha$) is associated with the $\kappa$ parameter via $\alpha =
\frac{1-p}{2} = \frac{2-\kappa}{2}$ (\citealt{rybi}).

Fit formulas for the synchrotron emissivities and absorptivities for thermal
and $\kappa$ distribution functions, which are used in the radiative transfer
models, were taken from \cite{symphony}. 

To capture all of the emission at low and high frequencies, we need a 
field of view for our camera of $2000 ~\rg$ (the extent of the GRMHD simulation domain). 
The code calculates the total flux density at every frequency by calculating null geodesics 
and simultaneously performing radiative transport calculations \citep{raptor}. Therefore it has the same field of view at every 
frequency. In the case of a uniform camera grid, one needs a very high resolution to 
resolve the source at both low and high frequencies simultaneously. This is because the high-frequency emission 
originates mainly from near the event horizon, a region that is much smaller than the extended jet structures seen at lower frequencies. In order to resolve the horizon 
with a uniform camera at this large field of view, one needs resolutions of around $10.000^2$ pixels, 
increasing the runtime of the code significantly. To overcome this runtime issue, 
and to obtain converged SEDs, we implemented a polar logarithmic camera grid into {\tt RAPTOR}. We describe the camera 
grid in Appendix \ref{camera} and show that our SEDs are converging if we use $512$ 
pixels in $\log{(r)}$ and $256$ pixels in $\theta$. 

 \begin{figure*}
\centering \includegraphics[width=\textwidth]{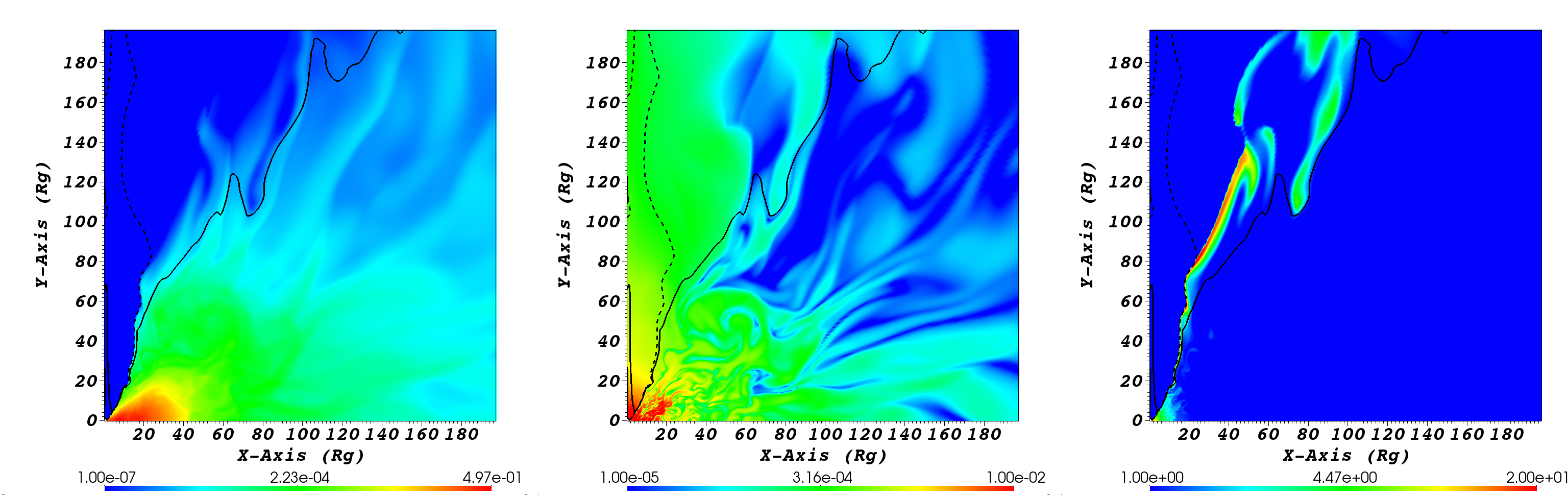}
        \caption{Color maps of the plasma density (left panel), the magnetic
          field strength (middle panel), and the dimensionless electron
          temperature at $t=4000 \,\tunit$ of the simulation. We show the inner
          regions of the upper half of the simulation domain. The gas density
          and magnetic field strength are given in the simulation code
          units. The electron temperature is computed using Eq.~\ref{Eq:Te}
          with $R_{\rm low} = 1$ and $R_{\rm high}=25$ (right panel). Each
          panel also shows contours of constant $\sigma=1$ (dashed contour)
          and ${\rm Be} = 1.02$ (solid contour). Regions where ${\rm Be}>1.02$ are outflowing. }\label{grmhd-pseudocolor}
  \end{figure*}
  \begin{figure*}
    \includegraphics[width=0.31\textwidth]{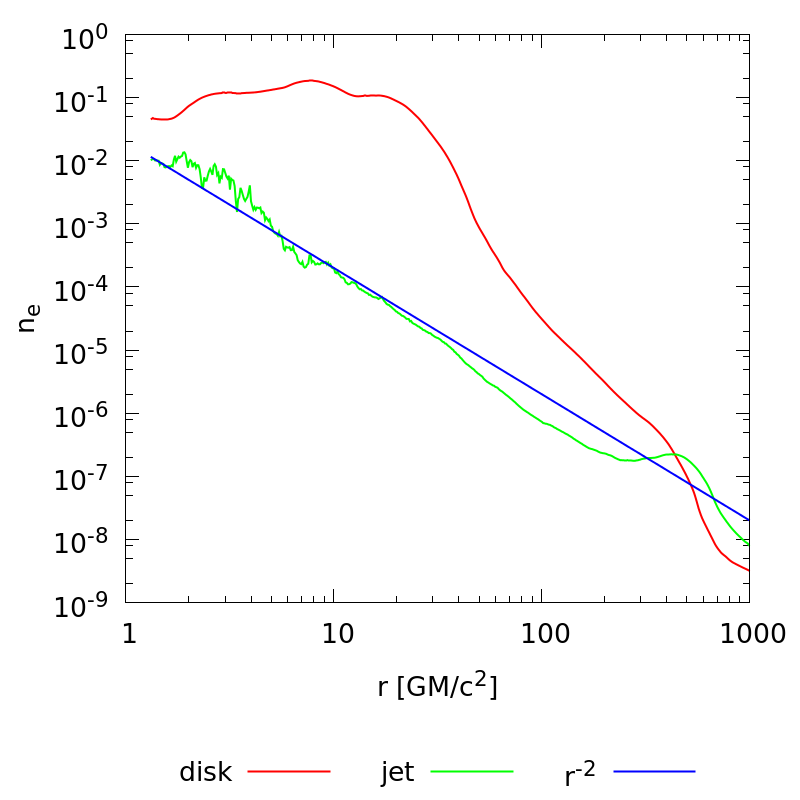}
    \includegraphics[width=0.31\textwidth]{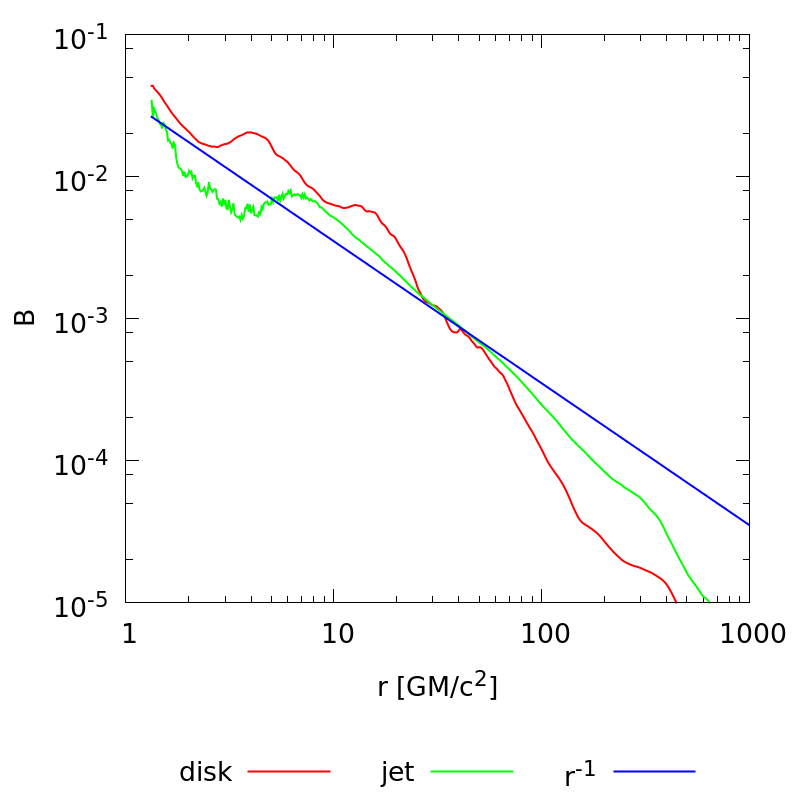}
    \includegraphics[width=0.31\textwidth]{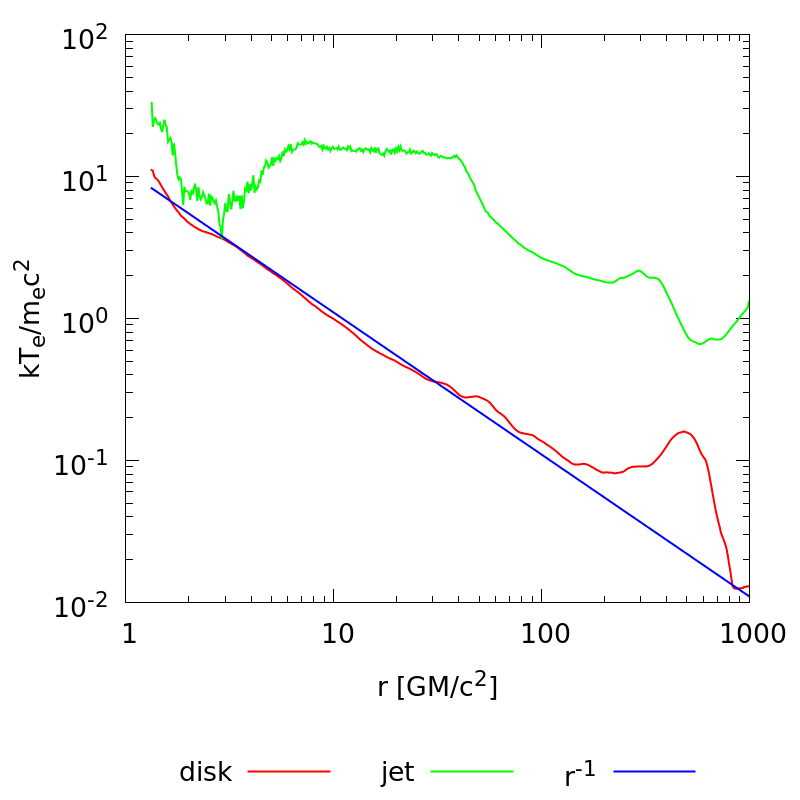}
        \caption{Radial profiles of the dimensionless number density (left
          panel), the magnetic field strength (middle panel), and the
          dimensionless electron temperature (right panel) shown in
          Figure~\ref{grmhd-pseudocolor}. Here, the simulation data is additionally time- and shell- averaged. Green and red lines
          correspond to plasma in the jet-sheath and accretion disk,
          respectively. Left and middle panels: blue lines represent the analytical model from BK79. Right panel: blue line represents the virial temperature profile.}\label{grmhd-lineprofiles}
        
\end{figure*}

\section{Results}\label{results}

\subsection{GRMHD jet structure}\label{jet-structure}

In Figure~\ref{grmhd-pseudocolor}, we show the
density, magnetic field strength, and electron temperature of a time slice of
the GRMHD simulation. In each panel, the color maps are overplotted with
contours of $\sigma=1$ and $Be=1.02$, which define the jet-sheath region where
electrons experience acceleration. The rightmost panel in
Figure~\ref{grmhd-pseudocolor} shows that a thin sheath of high-temperature
gas coincides with the unbound, outflowing matter.

Figure~\ref{grmhd-lineprofiles} displays time- and shell-averaged radial
profiles of $n_e$, $B$, and $T_e$ in the simulation. The radial profiles are
calculated using the following definition:
\begin{equation}
\langle q(r) \rangle = \frac{1}{\Delta t} \int_{t_{min}}^{t_{max}} \frac{\int
  \int_0^{2\pi} q(t,r,\theta,\phi) \sqrt{-g(r,\theta)} d\theta d\phi}{\int
  \int_0^{2\pi} \sqrt{-g(r,\theta)} d\theta d\phi} dt,
\end{equation}
where $t_{min}=3000 \,\tunit$ and $t_{max}=4000 \,\tunit$ and $\Delta t =
t_{max} - t_{min}$. The averaging uses 100 time slices of the GRMHD model, which corresponds to 5.47 hours.

Figure~\ref{grmhd-lineprofiles} shows that the radial profiles of the quantities
of interest follow power-laws.  We compare the radial dependencies of these
quantities to the Blandford-K{\"o}nigl jet model (\citealt{BKmodel}, hereafter
BK79, see also \citealt{falcke}), which is often invoked to explain the flat-spectrum radio cores observed
in the centers of active galactic nuclei. In the BK79 model, the jet density
decreases with radius as $n_e \propto r^{-2}$, the magnetic field strength as
$B \propto r^{-1}$, and the electron temperature 
$T_{\rm e}$ is constant.

We find that in our GRMHD simulations, the
electron temperatures are roughly constant up to $100 \,\rg$ in the jet-sheath.
A possible explanation for the temperature deviation from isothermality at
larger radii is that the initial torus is relatively small compared to the
size of the computational domain, and thus cannot collimate the jet at large radii;
without the pressure support of a large disk, the outflowing plasma in the jet decompresses adiabatically, which results in the observed temperature decrease. In the accretion disk,
the electron temperature follows a virial temperature profile, $T\propto r^{-1}$,
as expected. In our simulation, the radial profile of the magnetic field
strength does not resemble $B\propto r^{-1}$, but it is slightly steeper. It is
likely that the radial profile of the $B$ field strength along the jet in our numerical model is an artifact associated with the axisymmetric character of the simulation and the corresponding numerical difficulties. This difficulty arises because the turbulence in the magnetic field weakens due to the azimuthal symmetry of the 2D simulation. The radial structures of the inflows and outflows in the axisymmetric
GRMHD simulation carried out here are consistent with those presented in
similar axisymmetric models presented in \cite{noble2007} and in \cite{sgra}.

\subsection{SEDs and synchrotron maps of $\kappa$-jet models}\label{imaging}

In this section, we present spectral energy distributions
(SEDs) and radio, millimeter, and near-IR images of models where all electrons in the jet-sheath are described by the $\kappa$ distribution function.
The adopted free parameters of the model are listed in Table~\ref{tab:modelpar}.
We investigate how the model SEDs and images of the source change with the $\kappa$ parameter and with the observing angle. We measure the source sizes at various wavelengths using image moments \citep{imagemoments}. Aside from model-to-model comparison, we qualitatively compare the synthetic SEDs and source sizes to Sgr~A* observational data collected from the literature.

\begin{table}
\caption{List of parameters that are used in the radiative transfer
  simulations of Sgr~A*. The mass and distance of Sgr~A* are
  taken from \citet{gillesen}.}
  \label{tab:modelpar}
\centering
\begin{tabular}{cc} 
 \hline
 \hline
 Parameter  & value for Sgr~A* \\
 \hline
$i$ & $30^{\degree}$, $60^{\degree}$, $90^{\degree}$  \\ 
$D$ & 8 kpc \\ 
$M_{\rm BH}$ & $4.0\times 10^6\,\msun$  \\ 
$\mathcal{L}$ &$5.9 \times 10^{11}$ cm\\
$\mathcal{T}$ & 19.7 s\\
$\mathcal{M}$ & $10^{21}$ g \\
$\langle \dot{M} \rangle_t$ & $1.95 \times 10^{-8} \,\mdotu$\\
$\rho_0$ & $4.85 \times 10^{-15}$ [${\rm g \,cm^{-3}}$] \\
$n_0$ & $2.9 \times 10^{9}$ [${\rm \,cm^{-3}}$] \\
$B_0$ & $7409$ [Gauss] \\
$R_{\rm high}$ & 25 \\
$R_{\rm low}$ & 1  \\
$e^-$ accel. & jet sheath \\
$\kappa$ & 3,3.5,4,4.5,5\\
 \hline
\end{tabular}
\end{table}

\begin{figure*}
\centering
\includegraphics[width=\textwidth]{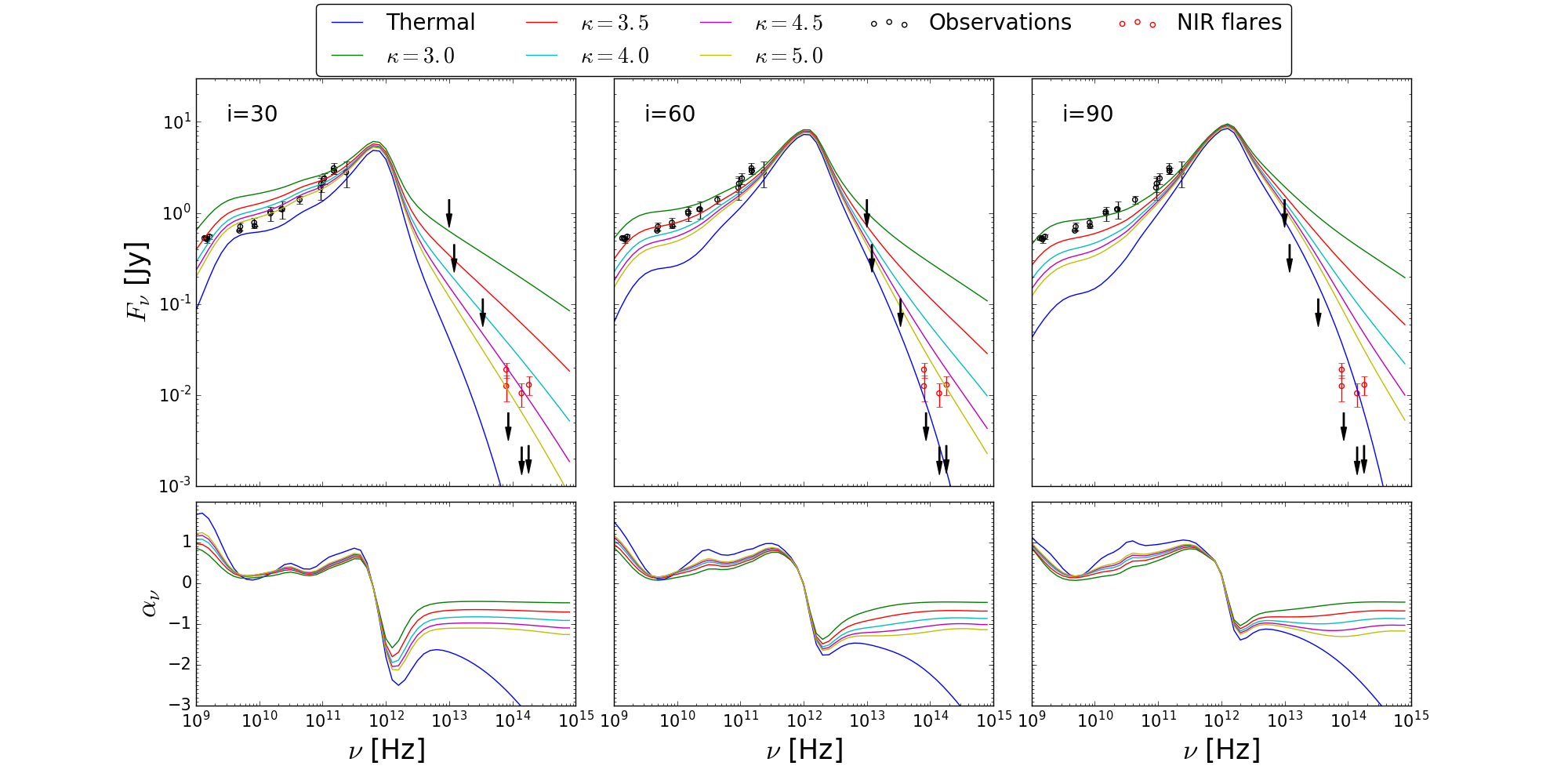}
\caption{SED overplotted with observational data (top) and spectral 
index (down) for thermal and $\kappa$ models for Sgr~A* at three different observing angles. 
Observational data from \citet{melia2001}, NIR flares from \citet{Genzel} and \cite{dodds-eden}.}\label{fig:spectra}
\end{figure*}

Figure~\ref{fig:spectra} (upper panels) shows model SEDs at three observing angles.  Since the observational data are
collected non-simultaneously, the model SEDs are time-averaged over
$\Delta t = 500 \,\tunit = 2.74$ hours. The best-fit model SED is for $\kappa=5$ 
and $i=30^\degree$. With these parameters, the model shows similar flux levels to the observed fluxes at radio frequencies
(\citealt{melia2001} and references therein), and is consistent with observational 
constraints at NIR frequencies (upper limits of NIR fluxes and flares). Unless the fact that we recover the correct flux values the spectral index of the best fit model is inconsistent with observational constraints ($\alpha \approx -0.7 \pm 0.3$ (\citealt{bremer2011})). Therefore a model where the electrons are distributed into a single $\kappa$ distribution function is, in the case of Sgr A*, ruled out, but can still be of interest for other sources, e.g., M87*. 

To decrease the spectral index, lower $\kappa$ values are needed. One could decrease the amount of NIR emission by decreasing the number of accelerated electrons at the jet launching point by having a mix of electrons in the $\kappa$ distribution function and a thermal distribution function. This idea is explored in Sect.~\ref{mix}.

Figure~\ref{fig:spectra} (lower panels) displays the spectral
slopes as a function of observing frequency.
Between $\nu=$10 and 230 GHz, the spectra for models with accelerated electrons have
flatter spectral slopes in comparison to the spectral slopes of the thermal model. At the NIR/optical frequencies, we observe a relation between the spectral index $\alpha$ and $\kappa$ given by $\alpha = \frac{2-\kappa}{2}$. This is expected behavior of optically-thin synchrotron emission because $\kappa = p + 1$ \citep{rybi}.

Our models demonstrate a strong dependence between the radio flux and the observer's viewing angle as predicted, e.g., by \citealt{falcke}. At lower inclinations, the jet points more towards the observer, the relativistic velocities inside the jet ($\gamma \beta \approx 5$), boost the emission to higher flux values. In the thermal model, we see lower flux values at radio frequencies
($\nu<10^{10} \,{\rm Hz}$) compared to fluxes obtained by \cite{sgra}. This is
because the authors of that paper assumed an isothermal jet up to $1000 \,\rg$, by setting the $T_{\rm e,jet}={\rm constant}$ inside the outflowing region of the simulation. When adding accelerated electrons in the jet, we
observe an increase in flux at the low and high-frequency sides of the
synchrotron bump. This is in agreement with results obtained by \cite{ozel}
and \cite{yuan}. As already mentioned in the introduction, these previous works used RIAF models, where the electrons are accelerated in the accretion disk. Our 
calculations show that it is also possible to recover the low-frequency "shoulder" and high-frequency "tail" by inserting the accelerated electrons in the jet outflow.

\begin{figure*}
\includegraphics[width=0.65\textwidth]{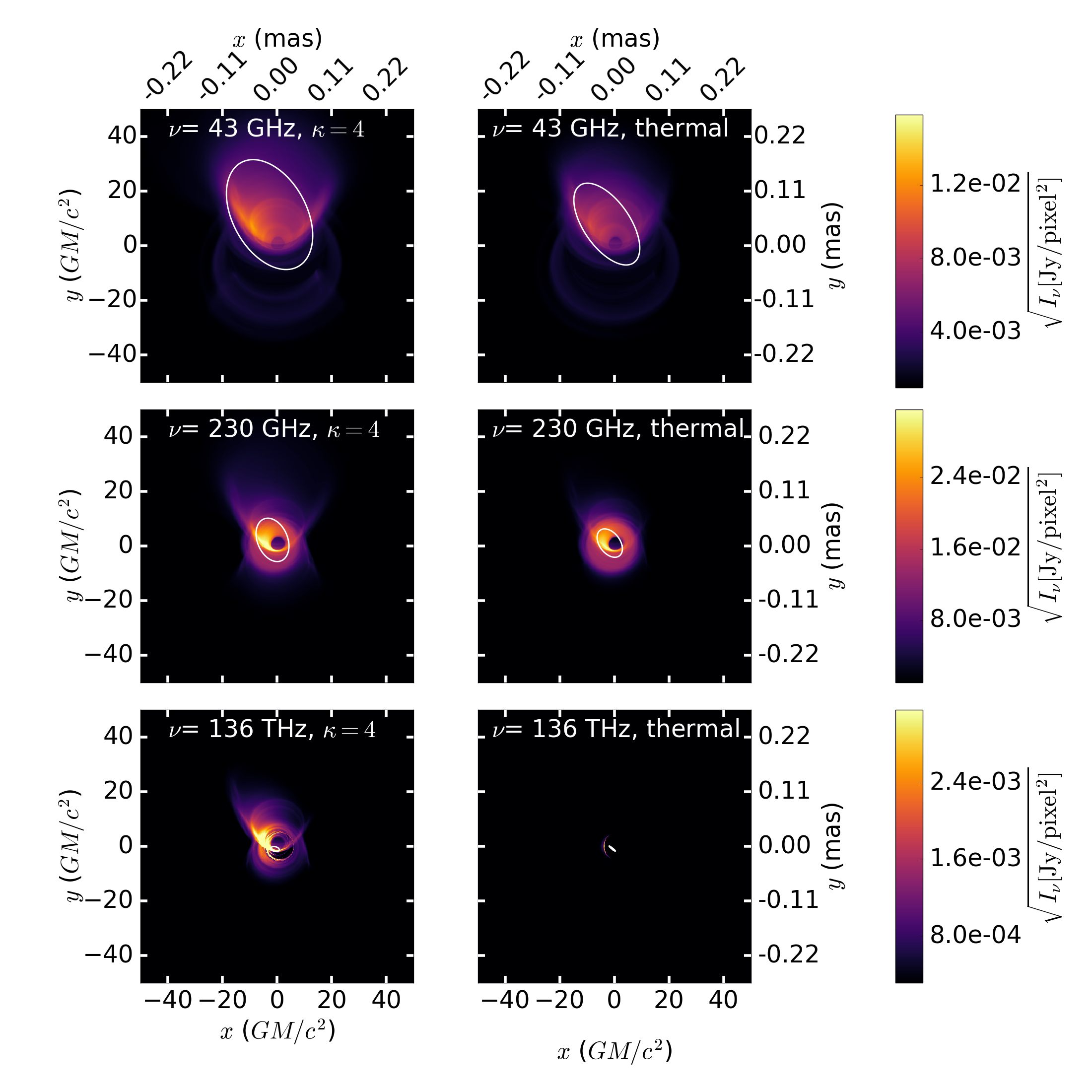} 
\includegraphics[width=0.33\textwidth]{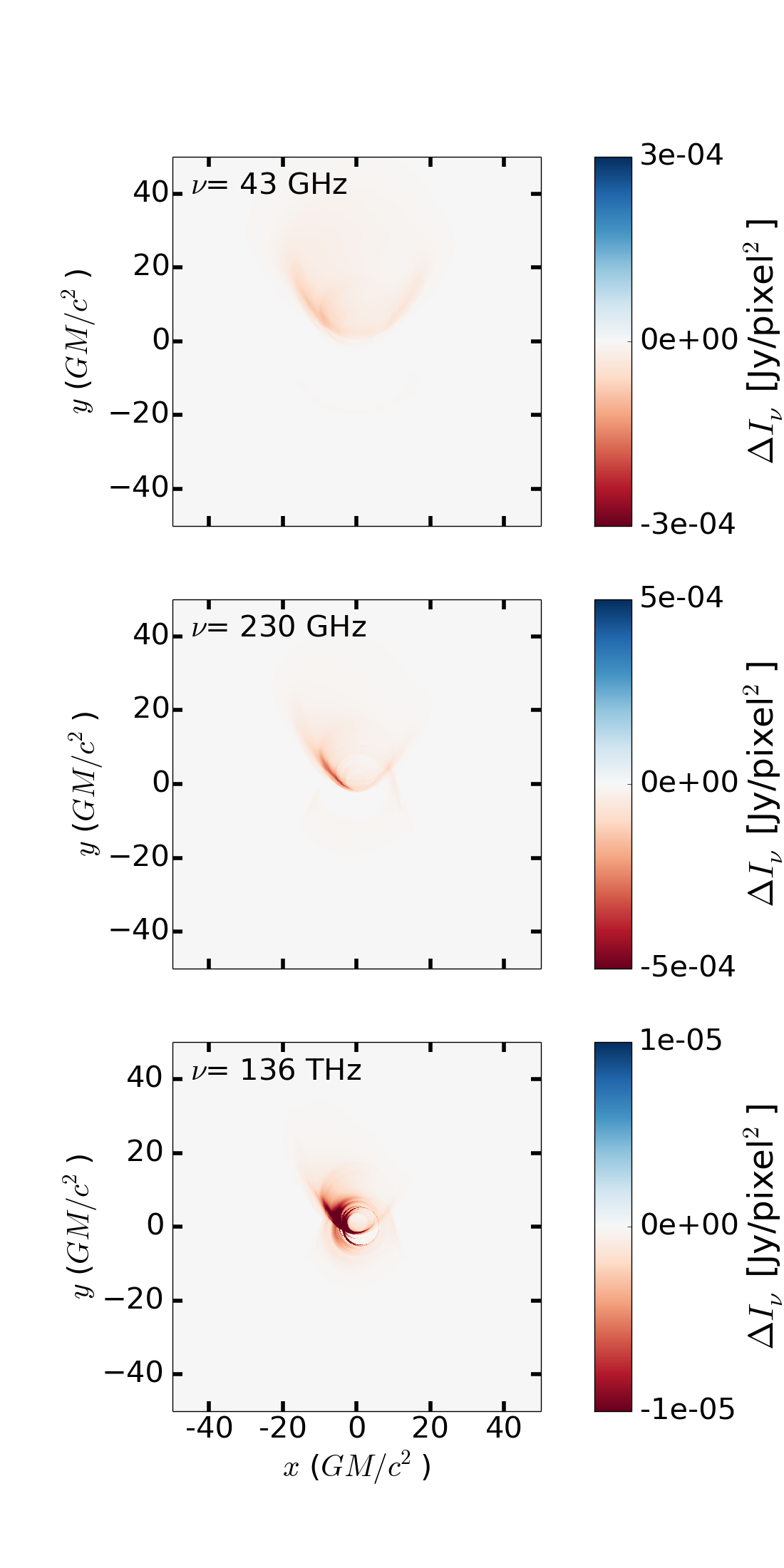}
\caption{Synthetic images for $\kappa$ models (left) and thermal models (middle) 
for Sgr~A* at three different frequencies for an observing angle  $i=30^\circ$ (with respect to the black hole spin axis). The 
spatial resolution and field of view of our camera is $0.2 \rg$ and 
$100 \rg$ respectively. Overplotted in white are the source sizes estimates, major and minor axis, and orientation of the ellipse, which are calculated by using image moments.
Right panels: the additional emission for Sgr~A* in the $\kappa=4$ model at three
  different frequencies. This emission is localized by subtracting the thermal
  synthetic images from the $\kappa=4$ images.
}\label{images_30}
\end{figure*}

\begin{figure*}
\includegraphics[width=0.65\textwidth]{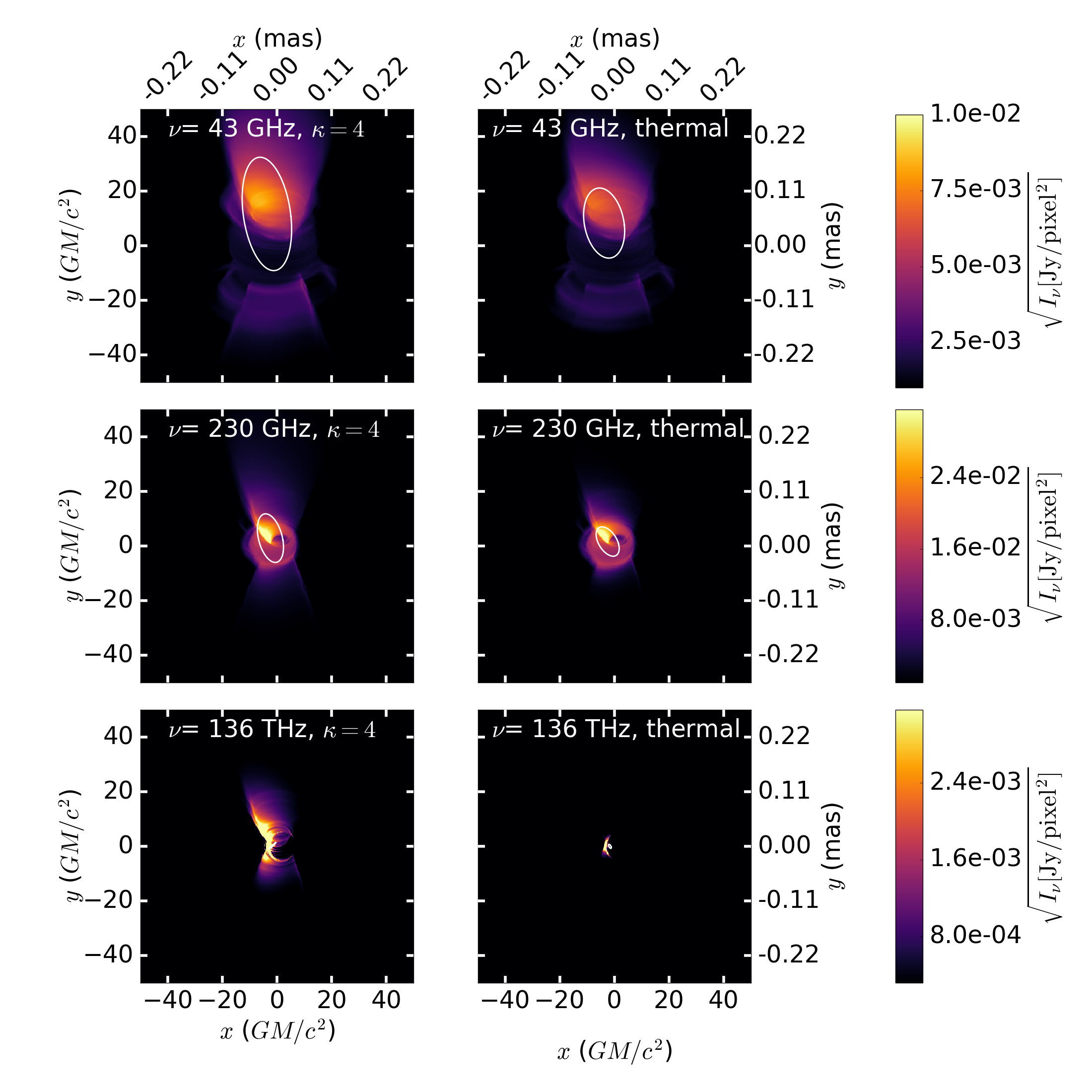} 
\includegraphics[width=0.33\textwidth]{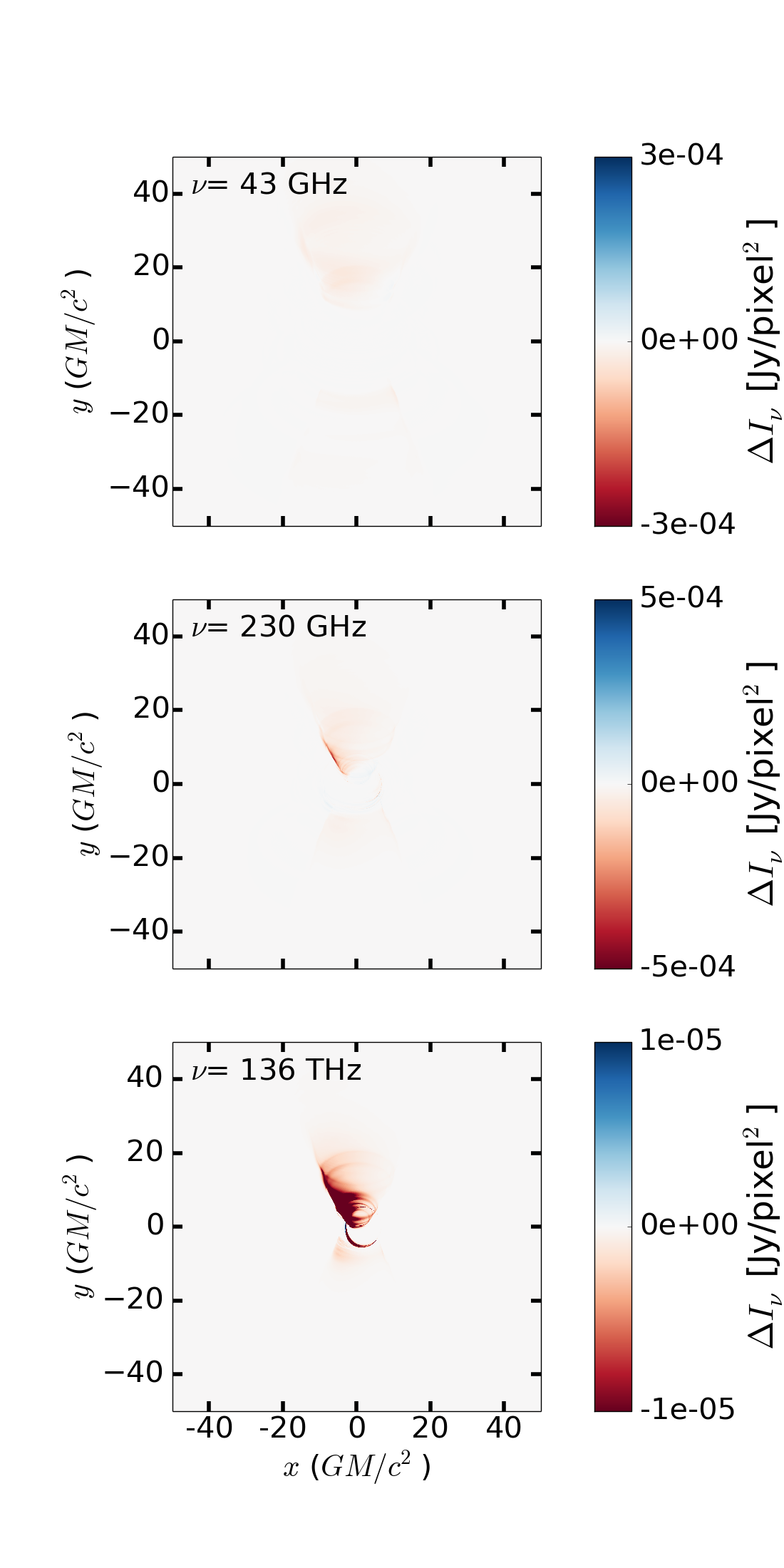}
\caption{As Fig.~\ref{images_30}, but for an observing angle of $i=60^\circ$.
}\label{images_60}
\end{figure*}

\begin{figure*}
\includegraphics[width=0.65\textwidth]{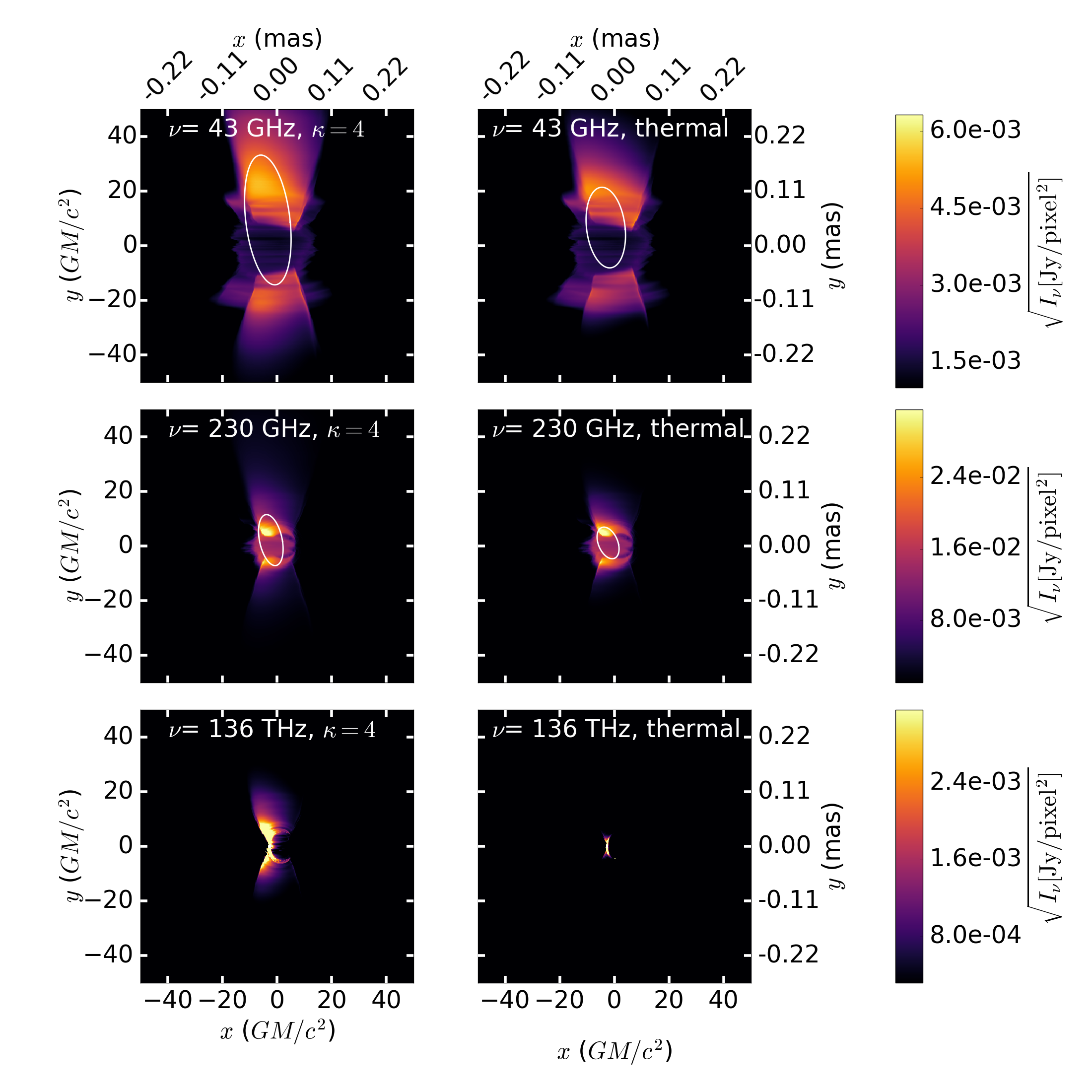} 
\includegraphics[width=0.33\textwidth]{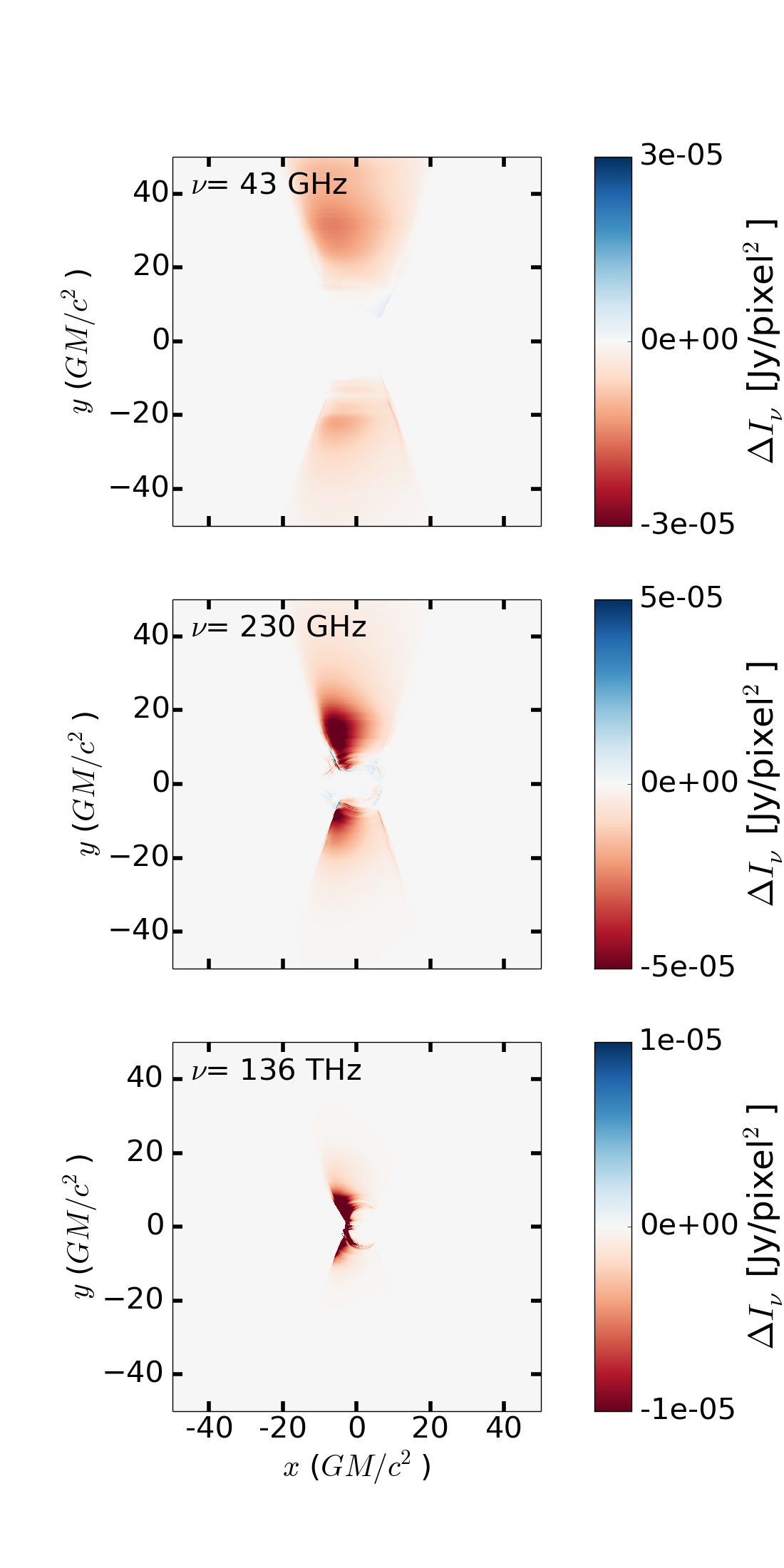}
\caption{As Fig.~\ref{images_30}, but for an observing angle of $i=90^\circ$.
}\label{images_90}
\end{figure*}

Figures~\ref{images_30},~\ref{images_60}, and ~\ref{images_90} 
show time-averaged radio, millimeter, and NIR
synthetic images of Sgr~A* for $\kappa=4$ (left panel) and for a relativistic thermal
electron distribution function (right panel) at three observing angles. The choice for $\kappa=4.0$ is arbitrary, and serves as an illustration of the difference between the $\kappa$ and thermal models. The synchrotron maps are overplotted with ellipses that represent the $FWHM$ of 
the major and minor axes of the source, as well as its orientation on the sky.
 In models with accelerated~ electrons, the jet is more elongated compared
to the models without electron accelerations. The difference in size is most noticeable 
in the NIR band. The extra emission produced by electrons in the high-energy tail is evident when subtracting the thermal model from the $\kappa=4$
model, as in the rightmost panels in Figs.~\ref{images_30},~\ref{images_60}, and~\ref{images_90}.

\begin{figure*}
\includegraphics[width=0.9\textwidth]{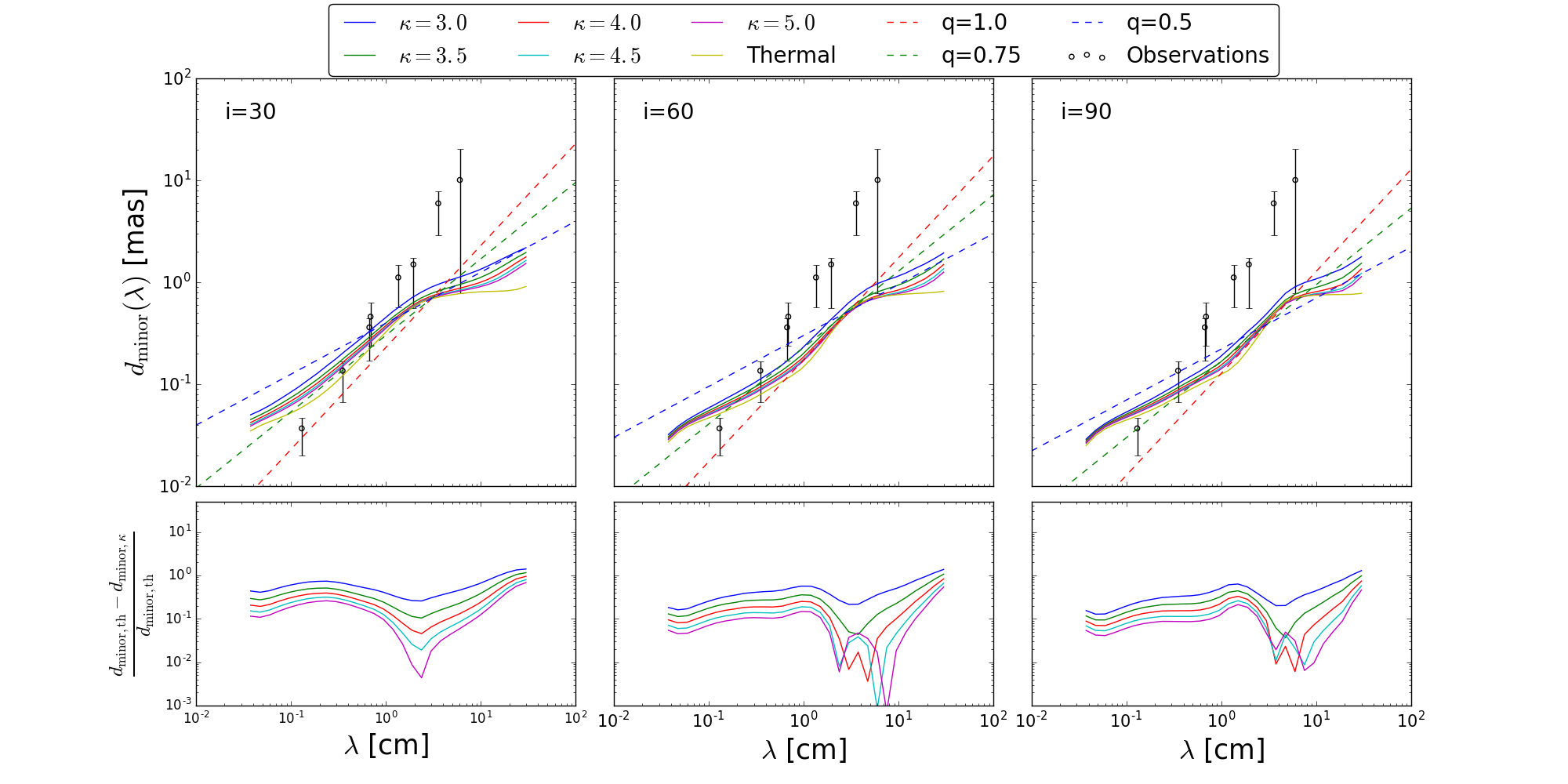}
\includegraphics[width=0.9\textwidth]{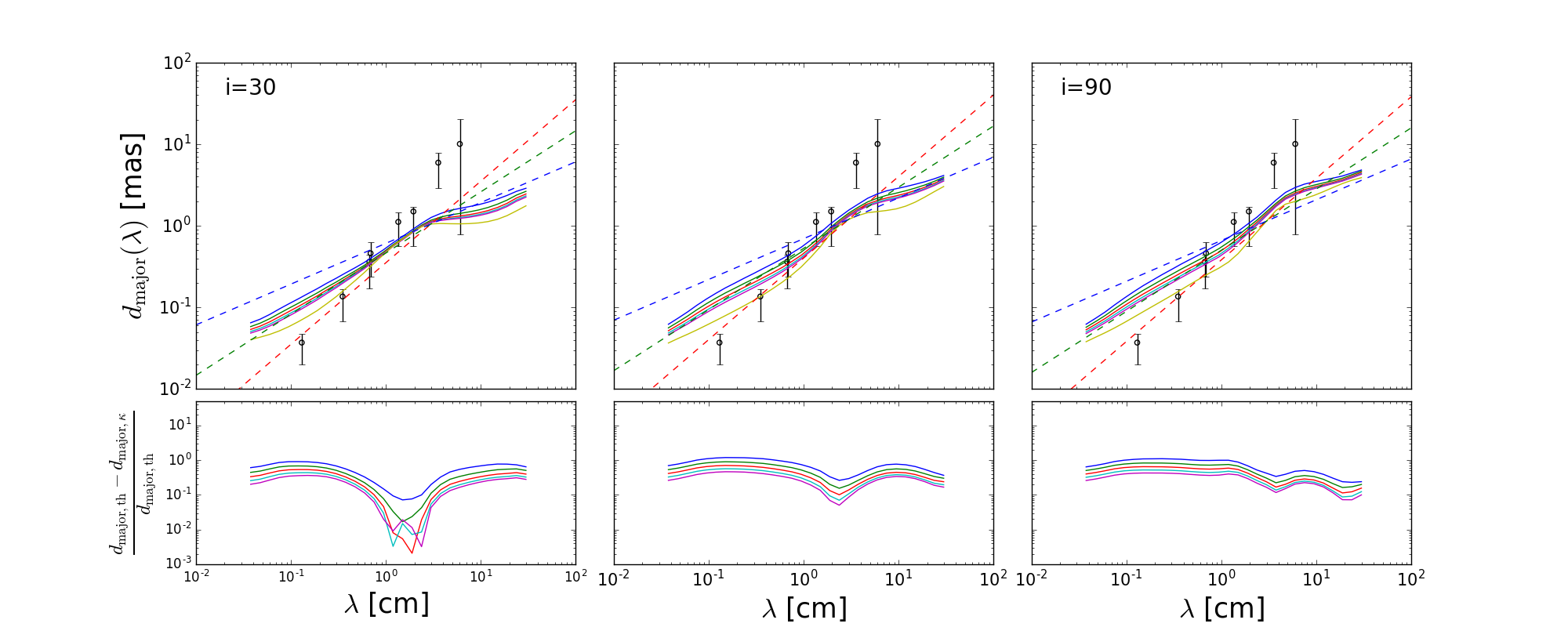}
\caption{First and third row: the major (first row) and minor (third row) 
axis sizes of the jet model as a function of observing wavelength for
 both thermal and $\kappa$ electrons together with 
 measured intrinsic sizes of Sgr~A* reported by 
 \protect\cite{bower2006} and \protect\cite{doeleman}.
 Dotted lines are $d(\lambda) =
 d_{th}(3 cm) \left(\frac{\lambda}{3 cm}\right)^q$ for three
 values of $q$. Second and fourth rows: 
 relative difference between $\kappa$-jet and thermal-jet size for the major 
 (second row) and minor axes (fourth row).}\label{sizesgra}
\end{figure*}

 Finally, Figure~\ref{sizesgra} compares the synthetic and observed (intrinsic, i.e.,
after subtraction of the scattering screen that 
 is detected towards the Sgr~A*) sizes of the emitting region for different $\kappa$ models at three observing angles in the optically thick part of the
spectrum. We plot both the major and minor axes of the source. 
Observationally, the source size follows a
power-law relationship as a function of $\lambda$: ${\rm size} \sim \lambda^q$, where $q=1$
(\citealt{bower2006}). We find that the major and minor size of the source model increase
with increasing observing wavelength $\lambda$, which is consistent with
observations.  At each wavelength, the model size increases with decreasing
$\kappa$ and decreases with increasing observing angle. Which $\kappa$ parameter 
recovers the observed size-$\lambda$ relationship best?  
The size of the minor axes is marginally consistent with the 1.3 mm observations in the case of the thermal and $\kappa$-jet models at $i=60^\degree$ and $i=90^\degree$ (see, two right top panels in Fig.~\ref{sizesgra}). All models produce jets
with sizes consistent with the error margins of the observations, but only down to 
$\lambda$=7 mm. At $\lambda<7$mm, all models overestimate the size of Sgr~A*. 
There are two possible explanations for this:

i) The axisymmetry of our simulations causes the appearance of ring-like structures in the flow which are also visible in the synthetic images. 
These ring-like structures are not present in 3D simulations, which may decrease the source size of the models. 

ii) Time-averaging the synthetic images results in concentrated 
emission that is smeared out over a larger volume as it propagates outwards, thereby 
increasing the measured model source sizes.

High-resolution, fully three-dimensional, radially extended GRMHD model of an accreting black hole with phenomenological prescriptions for the shape of non-thermal electron distribution function along the jet will be explored in a subsequent publication.


\subsection{Fitting the particle distribution function of Sgr~A*}\label{mix}

\begin{figure*}
\centering
\includegraphics[width=\textwidth]{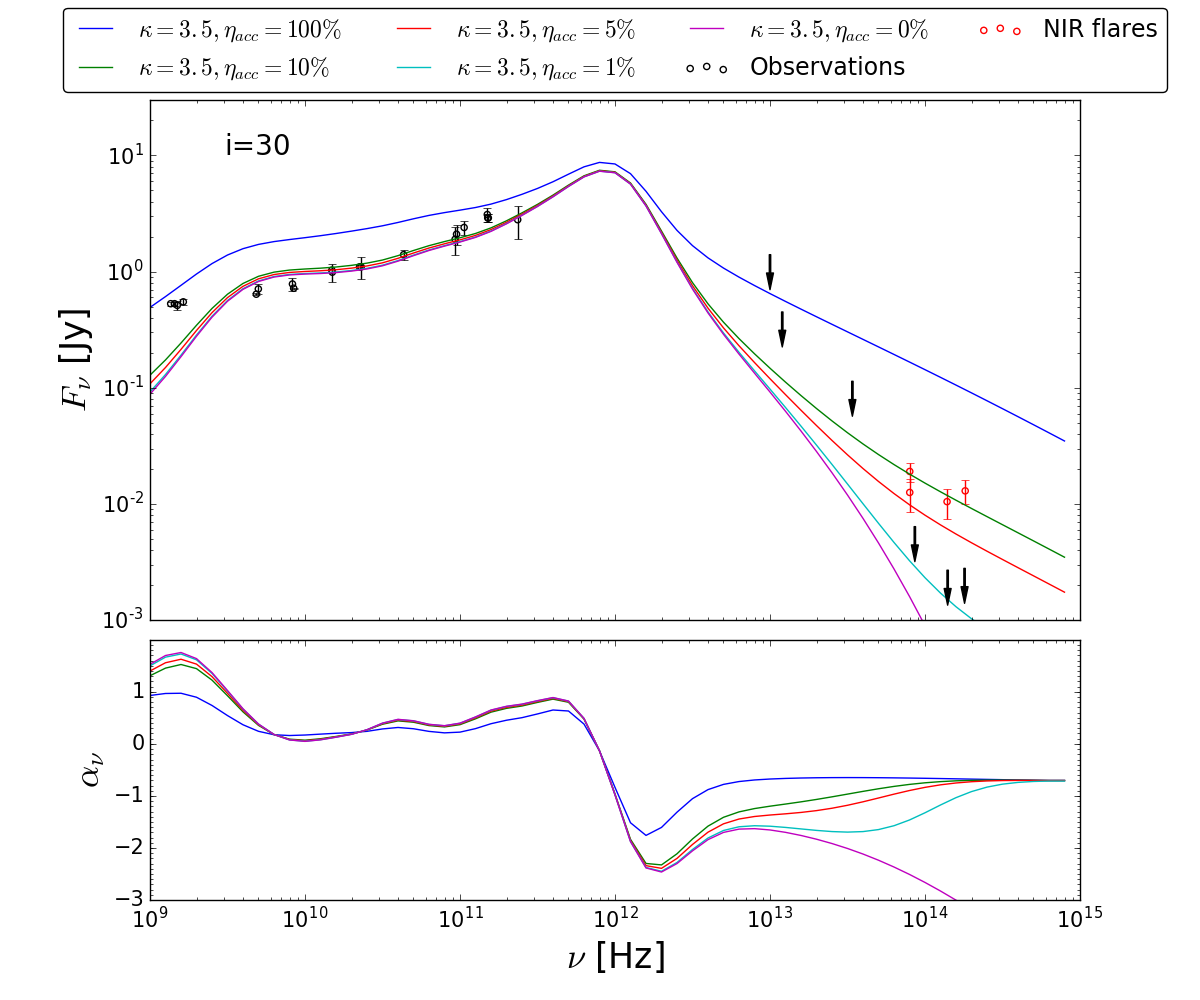}
\caption{SED overplotted with observational data (top) and spectral 
index (down) for various ratios of thermal and $\kappa$ models for Sgr~A* at an observing angle of $i=30^\degree$. 
Observational data from \citet{melia2001}, NIR flares from \citet{Genzel} and \cite{dodds-eden}.}\label{fig:spectra_mixed}
\end{figure*}

From observations of NIR flares, the spectral index of Sgr~A* is $\alpha \approx -0.7 \pm 0.3$ (\citealt{bremer2011}). This would result in a $\kappa$ value of 
$3.5$. The $\kappa=3.5$ models, however, overproduce the amount of flux in the NIR band. This is caused by injecting too many accelerated electrons in the jet sheath. In order to control the number of accelerated electrons, we use a superposition of a relativistic thermal and a $\kappa$ distribution in the jet sheath. The percentage of $\kappa$ distributed electrons is given by the free parameter $\eta_{\rm acc}$. The result of these fits for various values of $\eta_{\rm acc}$ can be seen in Figure \ref{fig:spectra_mixed}. 
The model where $\eta_{\rm acc} = 1\%$ fits the quiescence NIR observations, while a value of $\eta_{\rm acc} = 5 - 10\%$ fits the NIR flares. In both, the quiescence and flaring states, we recover the observed spectral index of $\alpha \approx -0.7 \pm 0.3$ (\citealt{bremer2011}). 

\section{Discussion}\label{discuss}

\subsection{SEDs of the jet launching zone as a function of electron distribution functions}

Various electron temperature models were used in the past to explain the
flat-to-inverted SEDs of radio jets.
In \cite{sgra} and \cite{sgra-3d}, an isothermal jet model was
introduced, the electron temperature was set to a constant value inside
outflowing regions of the accretion flow, and in the disk, the temperature
ratio was set to a constant value. 
In \cite{M87}, the temperature ratio between
the protons and electrons was described as a function of the plasma $\beta$
parameter. More recent work by \cite{ressler} showed that there is indeed a relation
between temperature and plasma $\beta$. 

In this work, we present a new set of models. We use the plasma $\beta$ 
prescription for the proton-to-electron temperature ratio in the disk, and 
we add accelerated electrons along the jet. The accelerated electrons are described by the $\kappa$ distribution function for electrons. With these
$\kappa$-jet models, we recover the flat-to-inverted SEDs reported by observers, while we relax the
assumption of an isothermal jet. The best fit pure $\kappa$ model fits the radio and NIR flux when $\kappa=5.0$ and $i=30$, but does not recover the spectral index of $\alpha = -0.7$ in the NIR. Therefore a mixed model of $\kappa$ distributed electrons and thermal electrons is favored.

If we use a mixed distribution (a superposition of a thermal and a $\kappa$ distribution) to describe the electrons inside the jet, instead of $\kappa$ only, we obtain a better fit to the observed SEDs. For NIR upper limits in quiescence, we obtain a fit with $\kappa = 3.5$, $\eta_{\rm acc}< 1\%$, and an observing angle $i = 30^\degree$, while for flaring states we obtain $\kappa = 3.5$, $\eta_{\rm acc}= 5-10\%$, and an observing angle $i = 30^\degree$. With these values we also recover the observed spectral index of $\alpha = -0.7 \pm 0.3$. By considering $\eta_{\rm acc}$ as a free parameter, we add an extra degree of freedom to our models. We think that it is justified to assume that only a subsection of the electrons will encounter the shock structures. All of our $100\%$ $\kappa$ jet models do not fit the spectral index of Sgr~A*; these models could, on the other hand, be valuable for different sources where the percentage of electrons that are accelerated could be large, e.g., M87*. Current GRMHD simulation cannot resolve shocks and are unable to capture the micro-physics of electron heating, future particle-in-cell simulations are necessary to fully understand the micro-physics involved. 

In the case of our best fit model where the percentage of electrons in the $\kappa$ distribution is $\eta_{\rm acc}=5\% - 10\%$, we can calculate the electrons acceleration efficiency as explained in \ref{ap:eff}. Our electrons acceleration efficiency $\eta$ for the mixed model results in $\eta = 0.06 - 0.12$, which is also similar to \cite{mao} and \cite{ball}, unless the fact that \cite{mao} and \cite{ball} insert the accelerated electrons in different regions. 

The best fit viewing angle is inconsistent with earlier papers like \cite{markoff2007}, where an inclination of $i>75^\degree$ is favored, and \citealt{broderick2009} and \cite{dexter2010}, who report an inclination of $i=50^\degree$. The inconsistency arises because lower viewing angles are necessary to fit the radio frequencies. This is a consequence of the relaxation of an isothermal jet since \citealt{sgra}  obtained fits with higher viewing angles. In our $\kappa$-jet models, inclinations higher than $\approx 60^\degree$ are excluded.

BK79 introduced an analytical model of a jet to describe nearly-flat spectra
radio cores of galaxies. Similar work was done by \cite{falcke}, who
showed a strong connection between the disk and the jet to explain radiative
properties of accreting black holes, such as radio luminosity and source size. It was assumed that the emitting electrons were in a power-law distribution; we
repeated these calculations in Appendix \ref{analyticalmodel}, but using the
$\kappa$ distribution function instead. We find a strong correlation between the source
radio-flux as a function of the $\kappa$ parameter, which decreases with increasing 
$\kappa$ values as can be seen in Figure~\ref{lnu}. We recover a radio flux that is independent of $\nu$, which is in agreement with both the BK79 model and \cite{falcke}.

The difference between the $\kappa$ and thermal models are relatively large at low- (radio) and high-frequency (optical/NIR) emission compared to the mm-wavelengths. The mm-emission is produced close to the disk, and the relatively small difference in flux density between the $\kappa$ and thermal models shows that the mm-emission is produced by the thermal electrons.

\subsection{Intrinsic size of the $\kappa$-jet model as a function of $\lambda$}

The synthetic radio images clearly show a more extended jet structure for 
Sgr~A* when emission from accelerated electrons is included in the
outflows. By
adding accelerated electrons, the energy in the population increases. In this circumstance, the more energetic electrons emit photons at higher frequencies compared to their
thermal counterparts. In general, for a given radiation frequency, the electrons radiate in different regions of the jet; the further away from the black hole one looks, the lower amount of emission is (since the magnetic field strength and the number density decay with increasing radius). When we accelerate the electrons in the jet, the energy available for emitting radiation increases. This results in a larger contribution to the total emission at larger radii compared to the thermal case, and hence in a more extended source. The observed and modeled core size follow a size-wavelength dependency $FWHM \propto \lambda^q$. In all $\kappa$-jet models, $q \lesssim 1$ for $\lambda > 3 \ {\rm cm}$. Our results can be understood in terms
of a simple model presented in Appendix~\ref{analyticalmodel}, where we
derive an analytical
expression that explains the source size as a function of $\kappa$. 

\section{Conclusion}
We analyzed the radial structure of jets produced in two-dimensional GRMHD simulations of an accreting black hole. Our simulations show a clear, thin jet-sheath region that follows the BK79 jet model in the inner~$100 \rg$, consistent with previous findings. 
The effects of various initial and boundary
conditions on the thermal structure of the radially 
extended jets should be
investigated in fully 3D models, because in 2D, the turbulence weakens due to azimuthal symmetry, and the
accretion rate, and hence the outflow rate, decrease over time.

We analyzed the impact of the particle acceleration in the jet-sheath on the observed radio flux and the jet emission region size. Our numerical results are confirmed by a simple semi-analytic jet model. We show that both, the radio flux and the size of the emitting region in the jet, increase with decreasing $\kappa$ parameter. At this time, all our models are too large compared to observational constraints from Sgr A* system, which
is again likely an artifact of axisymmetry.
However, our model easily recovers a nearly-flat radio SED of Sgr~A* while relaxing the assumption of a fully isothermal jet.
Our $\kappa$-jet model with $\kappa = 3.5$, $\eta_{\rm acc}= 1\%$ and observing angle $i = 30^\degree$ fits the Sgr~A* emission in quiescence. Additionally, our $\kappa$-jet model with $\kappa = 3.5$, $\eta_{\rm acc}=5\% - 10\%$, and observing angle $i = 30^\degree$, fits the observed fluxes of Sgr~A* when the source is in flaring state.

\section{\label{sec:level6}Acknowledgment}
This work was funded by the ERC Synergy Grant ``BlackHoleCam-Imaging the
Event Horizon of Black Holes'' (Grant 610058). We thank Jason Dexter for useful comments on the manuscript. This research has made use of NASA's Astrophysics Data System.
\nocite{*}

\bibliographystyle{aa}
\bibliography{references}

\appendix

\section{Polar logarithmic camera} \label{camera}

In order to resolve the emission profile at all frequencies, we adapted the camera of {\tt RAPTOR} to a logarithmic polar camera. In the case of a uniform camera grid, one needs a very high resolution to resolve both the low and high frequency in one SED. We therefore distribute our impact parameters (see \cite{raptor}) as follows:
\begin{eqnarray}
\alpha = r \cos(\theta),\\
\beta = r \sin(\theta),
\end{eqnarray}
where $r$ and $\theta$ are given by
\begin{eqnarray}
r = \exp\left(\log(r_{cam}) \frac{i}{N_r}\right) - 1,\\
\theta = \frac{2 \pi j}{N_\theta},
\end{eqnarray}
where $i,j$ are the pixel indices and $N_r,N_\theta$ are the number of pixels in $r$ and $\theta$ respectively.

\begin{figure}
\centering \includegraphics[width=0.5\textwidth]{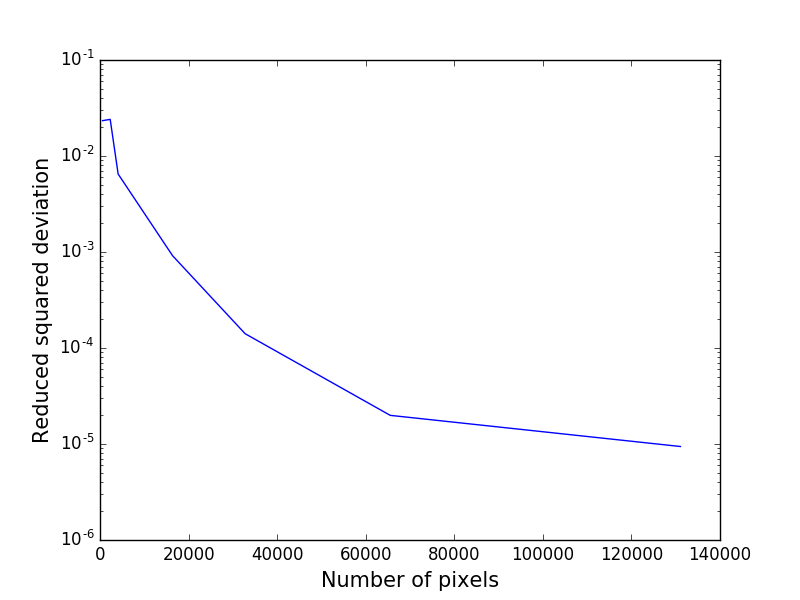}
\caption{ The relative difference between SEDs at different resolutions with respect to an SED at a resolution of $1024 \times 1024$. \label{error}}
\end{figure}

\begin{figure}
\centering \includegraphics[width=0.5\textwidth]{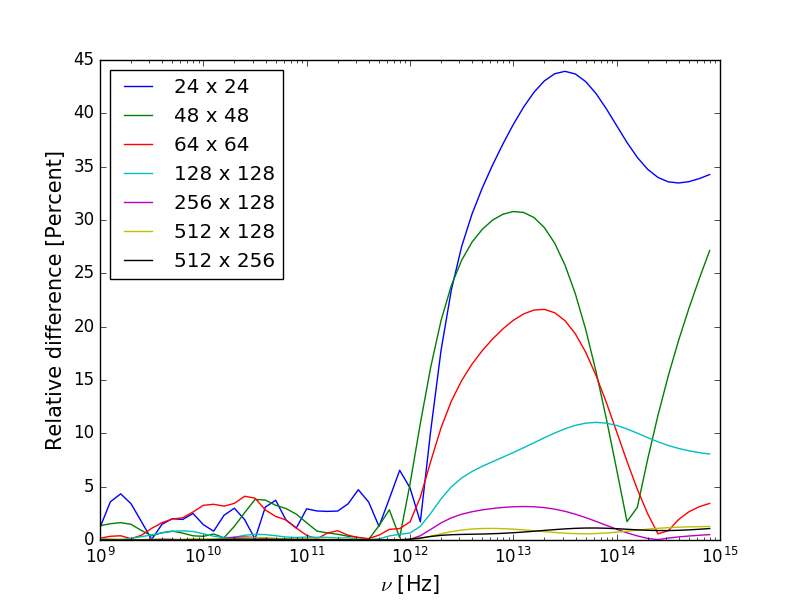}
\caption{ The relative difference between SEDs at different resolutions with respect to an SED at a resolution of $1024 \times 1024$. \label{relerror}}
\end{figure}

After the radiative transfer calculations, each intensity must be scaled by the size of the corresponding pixel; in polar coordinates, this surface element is given by $dA = r dr d\theta$, where
\begin{eqnarray}
dr = \frac{dr}{di} di = \frac{\log(r_{cam})}{N_r} \exp\left(\log(r_{cam}) \frac{i}{N_r}\right),\\
d\theta = \frac{d\theta}{dj} dj = \frac{2 \pi}{N_\theta}.
\end{eqnarray}
We calculated the convergence rate of this image grid by first calculating the square of the relative difference between the resolution under consideration and a high resolution polar grid of $1024 \times 1024$. We then sum this result over all frequencies, and divide this by the number of frequencies to calculate the reduced squared deviation. The deviation with respect to the high resolution run rapidly decreases several orders of magnitude with increasing resolution, as can be seen in Figure~\ref{error}.  
We show in Figure \ref{relerror} that the difference between our polar grid resolution of $512 \times 256$ is converged up to $\mathcal{O}(1)$ percent at all frequencies. It is evident from this image that, especially in oder to resolve the high frequency emission, one needs a high resolution image grid. 

\section{The size of a synchrotron photosphere as a function of $\kappa$ and
  $\lambda$} \label{analyticalmodel}

Here we calculate the source size and source radio luminosity of a relativistic magnetized jet as a function of
the $\kappa$ parameter. We follow the same approach as \cite{falcke} (FB95),
the only difference being that our distribution function is not a power-law
distribution function but the $\kappa$ distribution function. The jet in FB95
is assumed to be conically shaped, the opening angle of the jet being given
by
\begin{equation}
\phi \geq \arcsin \mathcal{M}^{-1},
\end{equation}
where $\mathcal{M}$ is the relativistic Mach number.

Our particle distribution function is given by:
\begin{equation}\label{kappafunction}
\frac{dn_{\rm {\rm e}\mid {\rm p}}}{d\gamma} = K_{{\rm e}\mid {\rm p}} \gamma \sqrt{\gamma^2 -1}
\left( 1 +
\frac{\gamma -1}{\kappa w}\right)^{-(\kappa + 1)},
\end{equation}
where $\gamma$ is the Lorentz factor, $w$ is a parameter that is equal to
the dimensionless temperature of the particles in the GRMHD simulation,
$\kappa$ is a free parameter that in the optically thin regime is related to
the power-law index $p$ by $\kappa=p+1$, and $K_{{\rm e}\mid {\rm p}}$ is a normalization
constant that determines the total amount of particles.

For simplicity, we assume that $w$ and $\kappa$ are the same for both the
electrons and protons. Similarly to FB95, we assume that there is an equipartition
between the magnetic energy $U_{B}$ and the energy of the particles $U_{\rm e+p}$
within a factor $k_{\rm e+p}$, such that
\begin{equation} \label{equipartition}
k_{\rm \rm e+p} \frac{B^2}{8\pi} = K_{\rm e} \int \gamma m_{\rm e} c^2 \frac{dn_{\rm e}}{d\gamma} d\gamma + K_{\rm p} \int \gamma
m_{\rm p} c^2  \frac{dn_{\rm p}}{d\gamma} d\gamma,
\end{equation} 
where $B$ is the magnetic field strength, $m_{\rm e}$ is the electron mass,
$c$ is the speed of light, and $m_{\rm p}$ is the mass of the proton.

Integrating Equation \ref{equipartition} results in
\begin{equation}
\begin{aligned}
k_{\rm e+p} \frac{B^2}{8\pi} = \frac{m_{\rm e} c^2 K_{\rm e} + m_{\rm p} c^2 K_{\rm p} } {{\Gamma  (\kappa+1)} }2^{\kappa-3} \kappa w \left(1-\frac{1}{\kappa  w}\right)^{-\kappa} \\  
\left(\frac{1}{\kappa w-1}\right)^{-\kappa}   \biggl(\kappa \Gamma \left(\frac{\kappa-3}{2}\right) \Gamma   \left(\frac{\kappa+1}{2}\right) \,  \\ 
_3F_2\left(\frac{\kappa}{2}-\frac{3}{2},\frac{\kappa}{2}+\frac{1}{2},\frac{\kappa}{2}+1;\frac{1}{2}, \frac{\kappa}{2};(\kappa w-1)^2\right) \\ 
-\kappa (\kappa+1) (\kappa w-1)   \Gamma \left(\frac{\kappa}{2}-1\right) \Gamma  \left(\frac{\kappa}{2}\right) \,  \\ 
_3F_2\left(\frac{\kappa}{2}-1,\frac{\kappa}{2}+1,\frac{\kappa}{2}+\frac{3}{2};\frac{3}{2},\frac{\kappa}{2}+\frac{1}{2};(\kappa w-1)^2\right)\biggr) \\
  = (m_{\rm e} c^2 K_{\rm e} + m_{\rm p} c^2 K_{\rm p})  \Lambda(\kappa,w),
\end{aligned} 
\end{equation}
where $\Gamma(x)$ is the Gamma function, $ _3F_2$ is the third
hypergeometrical function of the second kind, and the function
$\Lambda(\kappa,w)$ contains all dependencies on $\kappa$ and $w$.

We can then use this result to obtain an expression for the normalization
factor $K_{\rm e}$ to find
\begin{equation}
\begin{aligned}
K_{\rm e} = \frac{k_{\rm e+p}  B^2 }{8\pi \Lambda m_{\rm e} c^2 } ( 1 + \frac{m_{\rm p} K_{\rm p}}{m_{\rm e}
  K_{\rm e}}) \\ 
= \frac{k_{\rm e+p}  B^2 }{8\pi \Lambda m_{\rm e} c^2 } \mu_{p/e} \\
=\frac{  B^2 }{8\pi f m_{\rm e} c^2 },  
\end{aligned}
\end{equation}
where 
\begin{equation}
f = \frac{\Lambda \mu_{p/e}}{k_{\rm e+p}},
\end{equation}
and $\mu_{p/e}$ is the proton to electron ratio:
\begin{equation}
\mu_{p/e} = \left( 1 + \frac{m_{\rm p} K_{\rm p}}{m_{\rm e} K_{\rm e}}\right).
\end{equation}

With the normalization factors known, we can now calculate the number density of the
electrons by integrating equation \ref{kappafunction}:
\begin{equation}
\begin{aligned}
n_{{\rm e}\mid {\rm p}} = \frac{K_{{\rm e}\mid {\rm p}}}{\left(k^2-4\right) \Gamma (\kappa+1)}
2^{\kappa-2} \left(1-\frac{1}{\kappa w}\right)^{-\kappa-1} \\
\left((\kappa w-1)^2\right)^{\kappa/2} \biggl(4 \sqrt{(\kappa w-1)^2} \Gamma
\left(\frac{\kappa}{2}+2\right) \Gamma \left(\frac{\kappa}{2}\right) \\
   _2F_1\left(\frac{\kappa-2}{2},\frac{\kappa+2}{2};\frac{1}{2};(\kappa
w-1)^2\right)+2 \Gamma
   \left(\frac{\kappa-1}{2}\right) \\
   \Gamma \left(\frac{\kappa+3}{2}\right) \biggl(\kappa w (2-\kappa w) \,
   _2F_1\left(\frac{\kappa-1}{2},\frac{\kappa+3}{2};-\frac{1}{2};(\kappa
   w-1)^2\right)\\+(2 \kappa ((\kappa+1) w (\kappa
   w-2)+1)+1)
   \,\\ _2F_1\left(\frac{\kappa-1}{2},\frac{\kappa+3}{2};\frac{1}{2};(\kappa
   w-1)^2\right)\biggr)\biggr) \\
   = K_{{\rm e}\mid {\rm p}} \Phi = \frac{  B^2 }{8\pi f m_{\rm e} c^2 } \Phi(\kappa,w),
\end{aligned}
\end{equation}
where $_2F_1$ is the second hypergeometrical function of the first kind, and
$\Phi(\kappa,w)$ again contains all dependencies on $\kappa$ and $w$.\\

Similarly to FB95, we can define a ratio between the total number density and the
electron number density as
\begin{equation}
x_e = \frac{n_e}{n_{tot}},
\end{equation}
and a modified ratio as
\begin{equation}
x_e^{'} = \frac{x_e}{ \Phi(\kappa,w)}, 
\end{equation}
such that
\begin{equation} \label{eq:ntot}
n_{tot} = \frac{  B^2 }{8\pi f x_e^{'}  m_{\rm e} c^2 }.
\end{equation}
The mass supply of the jet is in FB95 given as a fraction of the mass supply
of the disk, which results in
\begin{equation}
\dot{M}_{jet} = q_m \dot{M}_{disk} = \gamma_j \beta_j c n_{tot} m_{\rm p} \pi
(r_{nozz} R_g)^2,
\end{equation}
where $q_m$ is the matter fraction of the outflowing matter, $\gamma_j$ is the
Lorentz factor of the bulk of the jet, $\beta_j$ is the bulk velocity of the
jet, and $r_{nozz}$ is the width of the jet nozzle.

We can use this expression to calculate $n_{tot}$, which is given by
\begin{equation}
n_{tot} = \frac{q_m \dot{M}_{disk}}{\gamma_j \beta_j c m_{\rm p} \pi (r_{nozz}
  R_g)^2}.
\end{equation}
We now use the result of equation \ref{eq:ntot} to find, for $B_{nozz}$,
\begin{equation}
B_{nozz} = \sqrt{\frac{8 q_m \dot{M}_{disk} m_{\rm e} c x_e^{'} f}{\gamma_j \beta_j
    m_{\rm p}}}.
\end{equation}
We have now obtained all initial conditions for the jet at the nozzle.

We use the same function for $r_{jet}(z_{jet})$ as FB95, which is given by
\begin{equation}
r_{jet}(z_{jet}) = r_{nozz} + (z_{jet} - z_{nozz})/\mathcal{M}, 
\end{equation}
where $r_{nozz}$ is the size of the nozzle of the jet, and $\mathcal{M}$ is
the Mach number. This relation asymptotically approaches
\begin{equation}
r_{jet}(z_{jet}) = z_{jet}/\mathcal{M},  
\end{equation}
if $z_{jet} \gg z_{nozz}$.

Conservation of mass and magnetic energy results in expressions for $B$ and $n_e$ as a function of radius given by
\begin{eqnarray}
B = B_{nozz} \mathcal{M}/z_{jet},\\
n_e = n_{e,nozz} / z_{jet}^{2}.
\end{eqnarray}
The internal energy also follows $z_{jet}^-2$, resulting in an isothermal
jet.

According to FB95, the Mach number is given by
\begin{equation}
\mathcal{M} = \frac{\gamma_j \beta_j}{\beta_s},
\end{equation}
which is the ratio between the proper flow speed and the sound speed of the jet. The sound speed is given by
\begin{equation}
\beta_s = \sqrt{u_{j0} (\Gamma^2 - \Gamma)f x^{'}_e m_{\rm e}/m_{\rm p} }.
\end{equation}

The optical depth of a conical jet is given by
\begin{equation}
\tau = 2 r_{jet}R_g \alpha_\kappa /\sin(i).  
\end{equation}
The location at which $\tau = 1$ is the point where the jet becomes
Synchrotron self-absorbed; this is a measure of the size of the
jet. To find this distance $Z_{ssa}$, we use the absorptivity based on the
$\kappa$ distribution function, where we assume that we are in the low
frequency limit;
\begin{multline}
\alpha_\kappa = \frac{n_e}{B sin(\theta)} X^{-5/3}_\kappa 3^{1/6}
\frac{10}{41} \frac{(2 \pi)^2}{(w \kappa)^{16/3 -
    \kappa}}\\ \frac{(\kappa-2)(\kappa-1)\kappa}{3\kappa-1} 
\Gamma\left(\frac{5}{3}\right)  \;_{2}F_{1}\left(\kappa - \frac{1}{3}, \kappa
+ 1, \kappa + \frac{2}{3}, - \kappa w \right) \\ = \frac{n_e}{B sin(\theta)}
X^{-5/3}_\kappa \chi(\kappa,w),
\end{multline}
where 
\begin{eqnarray}
X_\kappa = \frac{\nu}{(w\kappa)^2 \sin(\theta) \nu_c},\\
\nu_c = \frac{eB}{2\pi mc},
\end{eqnarray}
and $\chi(\kappa,w)$ again contains only dependencies on $\kappa$ and $w$.\\

Inserting $B(z_{jet})$, $n_e(z_{jet})$ and solving for $Z_{ssa}$ results in
\begin{equation}
Z_{ssa} = \left(\frac{\mathcal{M} \sin(i)}{2}\right)^{-3/5} n_{nozz}^{3/5}
B_{nozz}^{2/5} \nu^{-1} \sin(\theta)^{2/5} \left(\frac{m_{\rm e} c^2 }{ e
  (w\kappa)^2}\right)^{-1} \chi ^{3/5}.
\end{equation}

\begin{figure}
\centering \includegraphics[width=0.45\textwidth]{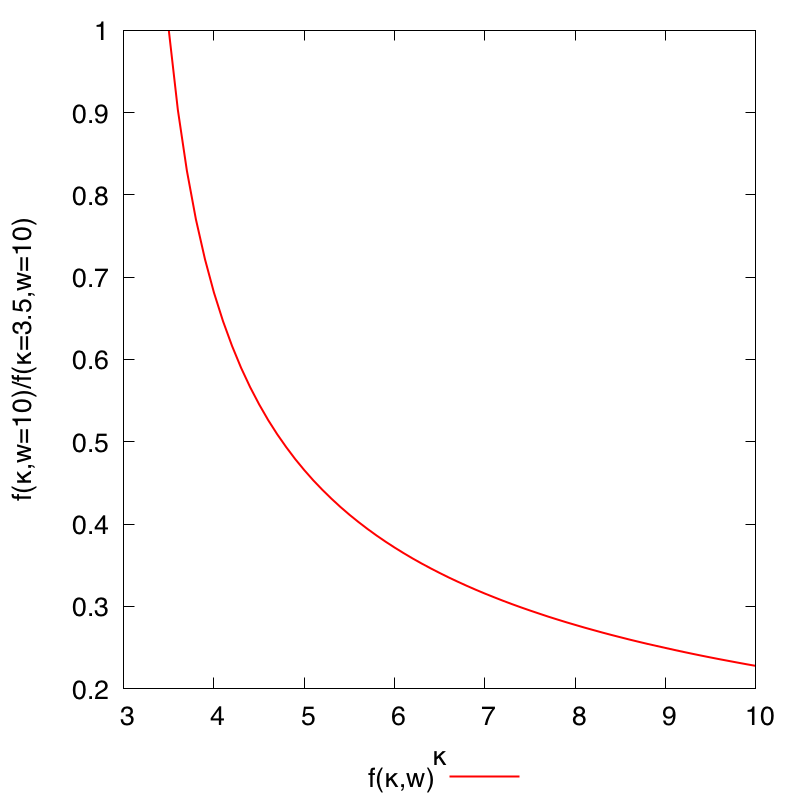}
\caption{The source size at a given frequency as a function of
  $\kappa$, where the size is defined as the radius at which the
  source transitions from optically thick to thin.\label{zssa}}
\end{figure}

We find a relation for the size as a function of frequency given by $Z_{ssa}\propto
\frac{1}{\nu}$. As a function of wavelength this results in $Z_{ssa}\propto
\lambda$ which is in agreement with recent observations of Sgr A*
(\citealt{doeleman} and \cite{bower2006}). Inserting $\mathcal{M}$, $B_{nozz}$ and $n_{nozz}$ results
in a relation between the source size and the $\kappa$ parameter given by
\begin{equation}
Z_{ssa} \propto {(w\kappa)^2}{\chi^{3/5}}
\left(\frac{\Lambda}{\Phi}\right)^{4/5} = f(\kappa,w),
\end{equation}
where the function $f(\kappa,w)$ is plotted in figure \ref{zssa}.\\

We can now also calculate the total radio luminosity by integrating
\begin{equation}
L_\nu = \int_{Z_{nozz}}^{Z_{ssa}} \epsilon(z_j)_\kappa \pi (z_j/\mathcal{M})^2
dz_j,
\end{equation}
where we use that the emissivity is given by
\begin{equation}
\begin{aligned}
\epsilon(z_j) =  \frac{ n_e e^2 \nu_c \sin (\theta)}{c} X^{1/3}_\kappa
\frac{4\pi\Gamma
  \left(\kappa-\frac{4}{3}\right)}{3^{7/3}\Gamma(\kappa-2)}  \\  
= \frac{ n_e  e^2 \nu_c \sin \theta}{c} X^{1/3}_\kappa \Theta(\kappa,w),
\end{aligned}
\end{equation}
and $\Theta(\kappa,w)$ again only depends on $\kappa$ and $w$.

If we assume that we can neglect the lower boundary of the integral (where
$\tau \gg 1$), we obtain
\begin{equation}
L_\nu = \frac{12 \pi e^3}{mc^3} \sin^{-1/5}(\theta)
\biggl(\frac{\mathcal{M}\sin(i)}{2}\biggr)^{-1/5} n_{nozz}^{6/5}
B_{nozz}^{4/5} \chi^{3/5}.
\end{equation}

\begin{figure}
\centering \includegraphics[width=0.45\textwidth]{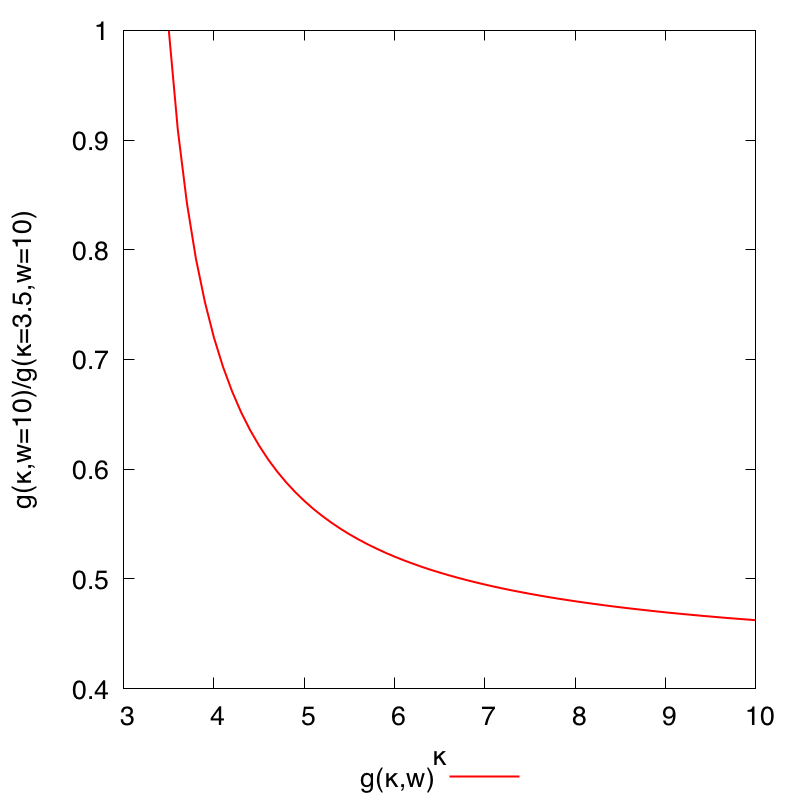}
\caption{ The radio flux in the optically thick part of the SED as a function of $\kappa$. \label{lnu}}
\end{figure}

The total radio flux is independent of frequency, as one would
expect. Inserting $\mathcal{M}$, $B_{nozz}$ and $n_{nozz}$ results in a relation between
$L_\nu$ and $\kappa$ given by
\begin{equation}
L_\nu \propto \left(\frac{\Lambda(\kappa,w)}{\Phi(\kappa,w)}\right)^{3/5}
\chi(\kappa,w)^{1/5} = g(\kappa,w).
\end{equation}
The resulting function $g(\kappa,w)$ is plotted in figure \ref{lnu}.

\section{Particle acceleration efficiency} \label{ap:eff}
\begin{figure*}
\centering \includegraphics[width=0.31\textwidth]{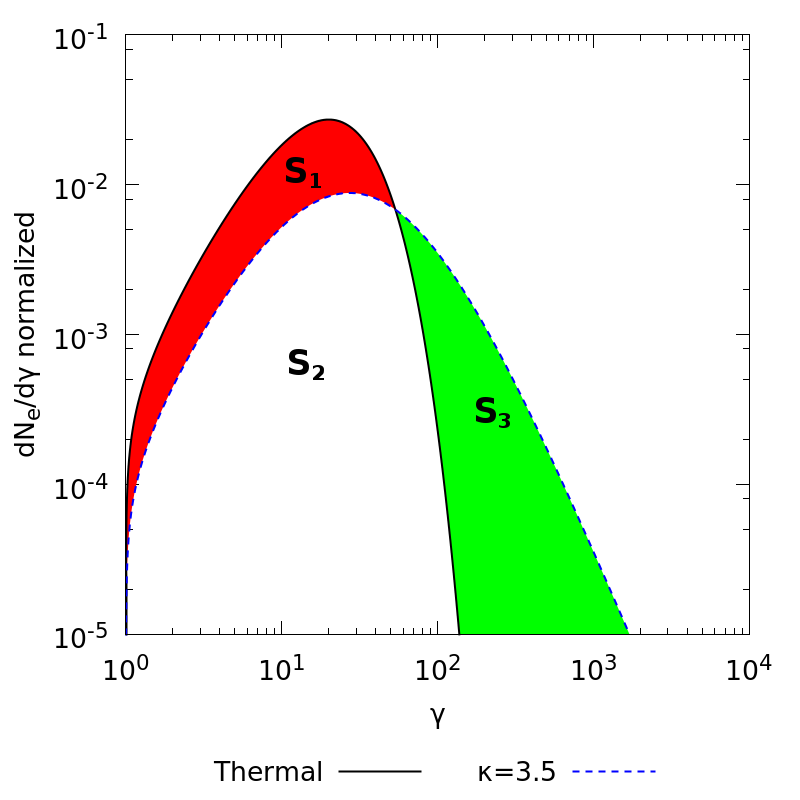}
\includegraphics[width=0.31\textwidth]{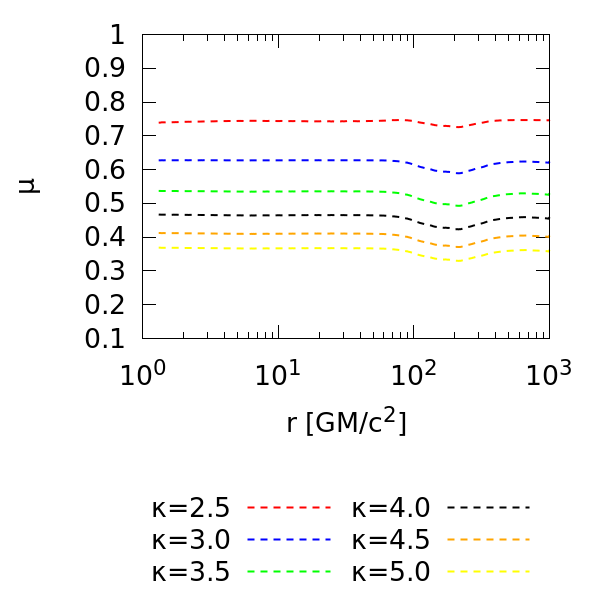}
\includegraphics[width=0.31\textwidth]{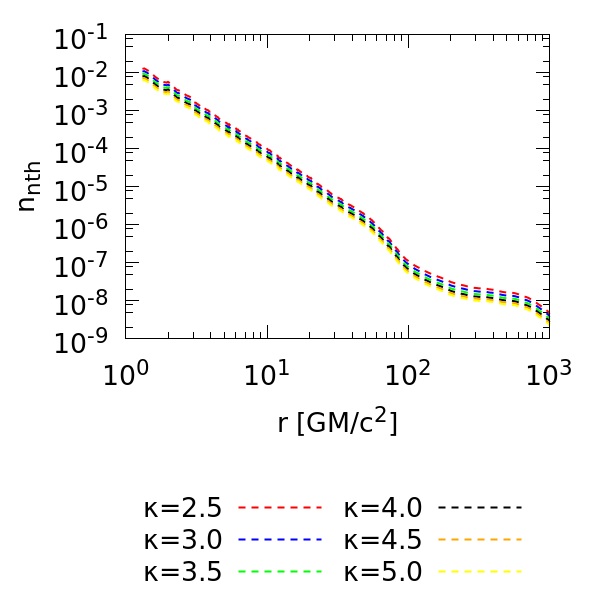}
\caption{Left panel: an illustration of the calculation of $\mu$ (Eq. \ref{mu}) for
  one value of $\kappa$.  Middle: ratio of particle number densities between the power law tail and the thermal core of the
  $\kappa$ distribution function along the jet. Right panel: Number density,
  in code units, of electrons in the power-law part of the $\kappa$
  distribution function. }\label{ratio}
\end{figure*}

What is the  electron acceleration efficiency in our pure $\kappa$-jet models as a function
of $\kappa$ parameter?  The $\kappa$ distribution function smoothly connects a
power-law to a thermal core; it is, therefore, difficult to define the electron
acceleration efficiency as was done in previous works. To compute the
efficiency, we introduce our modified acceleration efficiency $\eta_{mod}$,
which is defined as the ratio of the electrons that got shifted to higher
$\gamma$ values (the power-law part), compared to a purely thermal
distribution, and the electrons that experience only a small shift (the thermal
core). 
The thermal core and power-law part of the $\kappa$ distribution function are
defined in Figure~\ref{ratio} (left panel). If we compare a purely thermal distribution with a $\kappa$ distribution we can distinguish three different regions; $S_1$, $S_2$, and $S_3$. $S_1$ is the region where the thermal distribution is larger than the $\kappa$ distribution, $S_2$ is the region where they overlap, and $S_3$ is the region where the $\kappa$ distribution is larger. Since both distribution function are normalized to the same value, we know that $S_1$ and $S_3$ have to be equal in surface, and therefore, contribute an equal amount of electrons to the total amount of electrons. The consequence of this is that, when comparing a $\kappa$ model to the thermal case, the number of electrons in $S_1$ in the thermal models is shifted towards higher energies in the region $S_3$ in the case of the $\kappa$-jet models. The region $S_2$ is the number of electrons that are in the thermal core of the $\kappa$ distribution function. This enables us to quantify the number of electrons that shift to higher energies by integrating the difference between the thermal and $\kappa$ distribution up to the point $\gamma_{\rm max}$ where the two distribution functions are equal ($n_{\rm thermal}(\gamma_{max},\Theta_e) = n_\kappa(\gamma_{max},\Theta_e)$) 

Figure~\ref{ratio} (middle panel) shows the ratio between the electrons that get
shifted to the power-law tail and the total number density of electrons along
the jet as a function of the distance from the core for adopted values of the $\kappa$ parameter. Note that the particle number densities in the S1 and
S3 regions in Figure~\ref{ratio} (left panel) are equal, hence the fractional number
density of electrons in the power-law tail can be defined as:
\begin{equation}\label{mu}
\mu(r) = \frac{ \int_{1}^{\gamma_{max}} (n_{\rm
    thermal}(\gamma,\Theta_e) - n_\kappa(\gamma,\Theta_e)) \,
  d\gamma}{\int_1^\infty n_{\rm thermal}(\gamma,\Theta_e) d\gamma},
\end{equation}
where integration in the numerator is carried out between $1$ and
$\gamma_{max}$ and where the electron temperature $\Theta_e(r)$ is a function of radius as displayed in the rightmost panel in
Figure~\ref{grmhd-lineprofiles}. The fractional number density in the jet
region increases with decreasing $\kappa$ values, and is nearly constant as a
function of radius.

Figure~\ref{ratio} (right panel) displays the number density of electrons in
the $\kappa$ distribution function that occupy the power-law part of the
distribution function. Note that the particle number density is given here
in code units. To convert these values to [particles/cm$^3$], one has to
multiply with $n_0$, which is given for Sgr~A* in
Table~\ref{tab:modelpar}.

\begin{figure*}
\centering \includegraphics[width=0.31\textwidth]{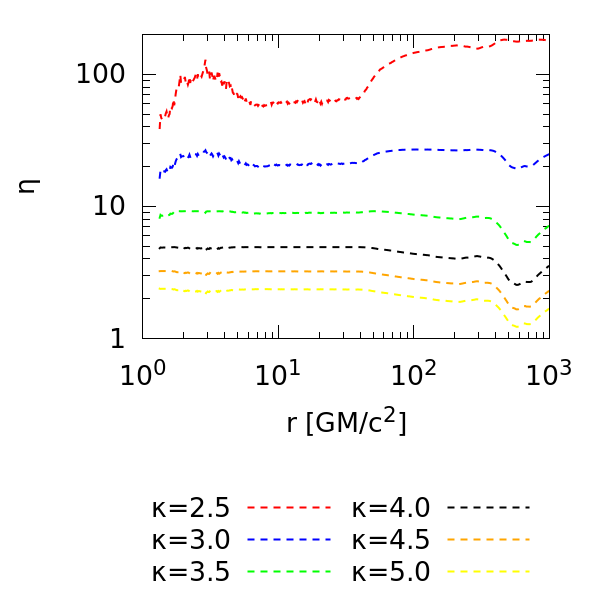}
\includegraphics[width=0.31\textwidth]{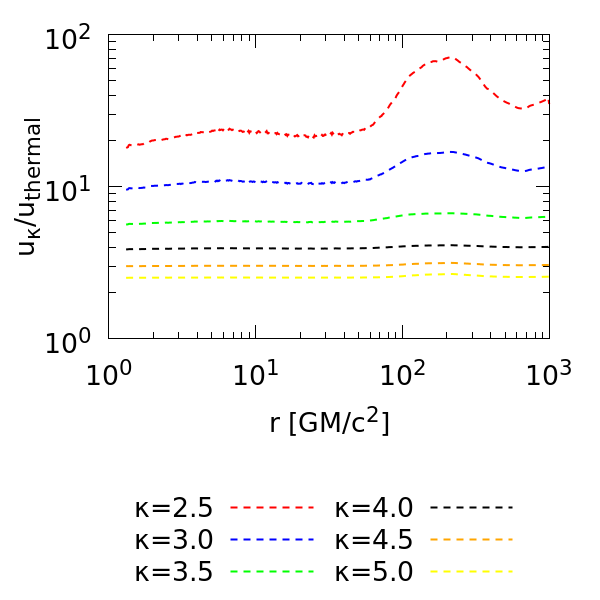}
\includegraphics[width=0.31\textwidth]{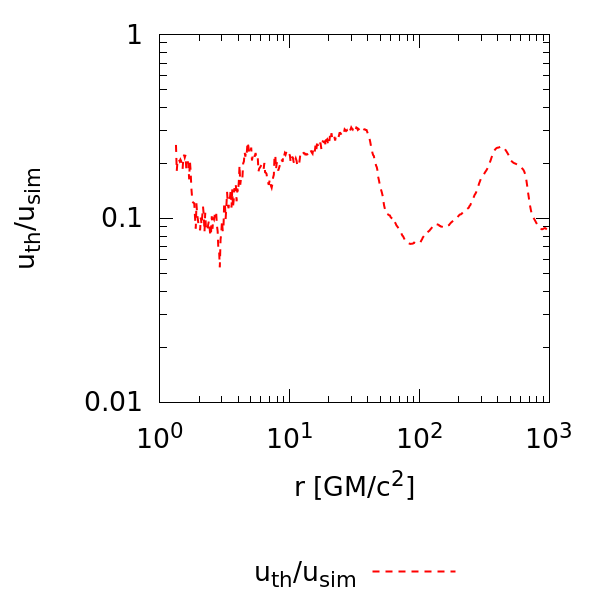}
\caption{Left panel: our modified energy efficiency $\eta_{mod}$. Middle:
  ratio of kinetic energy in $\kappa$ to a purely relativistic  Maxwell-J\"uttner  distribution
  function along the jet. The electron temperatures along the jet are shown in
  Figure~\ref{grmhd-lineprofiles} (right panel).  Right panel: ratio between
  the energy in the thermal distribution to the simulation
  energy.}\label{ratio2}
\end{figure*} 

Finally, we define the modified acceleration efficiency $\eta_{mod}$ as the
ratio of the total energy of the electrons in the S3 region to those in the S2 region, i.e. the ratio of energy in
the power-law tail to the energy in the quasi-thermal core:
\begin{equation}
\eta_{mod} = \frac{\langle u\rangle_{\kappa} (S3)}{\langle
  u\rangle_{\kappa}(S2)},
\end{equation}
where $\langle u\rangle_{\kappa}=\int (\gamma - 1) n_\kappa (\gamma,\Theta_e)
d\gamma$ is the kinetic energy of electrons integrated over the energy
distribution function. Figure~\ref{ratio2} (left panel) shows the modified
particle acceleration efficiency for various values of $\kappa$.

In Figure~\ref{ratio2} (middle panel), we compare the total energy in the $\kappa$ distribution
to the total energy in a purely relativistic thermal distribution function. The ratio of
$\kappa$ and thermal kinetic energies is given by
\begin{equation}
\frac{\langle u \rangle _{\kappa}(S2+S3)}{\langle u \rangle _{\rm thermal}} =
\frac {\int (\gamma - 1) n_\kappa (\gamma,\Theta_e) d\gamma}{\int (\gamma - 1)
  n_{\rm thermal} (\gamma,\Theta_e) d\gamma}.
\end{equation}
We find that for the smallest $\kappa$ parameter, the
energy within a radius of $100 \,\rg$ is about 20 times higher compared to a purely relativistic thermal distribution, and this increases to 70 times higher at larger
radii, where the temperature in the jet decreases, and can, therefore, explain the
increase in energy ratio. The Maxwell-J\"uttner distribution function narrows
for small values of $\Theta_e$, hence adding a power-law has a larger
relative effect on the energy content while the absolute difference is
smaller.

How much energy is in thermal electrons (and the $\kappa$ distribution function)
compared to the energy available in the simulation? The total energy in
thermal electrons is analytically given by (\citealt{Gammie1998})
\begin{equation}
u_{\rm thermal} = a (\Theta_e) N_{\rm thermal} m_{\rm e} c^2 \Theta_e,
\end{equation}
where
\begin{equation}
a(\Theta_e) \approx \frac{6+15\Theta_e}{4 + 5 \Theta_e}.
\end{equation}
Figure~\ref{ratio2} (right panel) shows the ratio $u_{\rm thermal}/u_{\rm
  sim}$, the energy in a thermal distribution to the energy available in the
simulations:
\begin{equation}
u_{\rm sim} = u_{\rm int} + B^2 / 2,
\end{equation}
where $u_{\rm int}$ is the internal energy density, and $B$ the magnetic field
strength. The value of this ratio is around $0.1 - 0.3$. We can therefore only 
use radiative transfer models that are self consistent, i.e. 
$\frac{u_\kappa}{u_{\rm thermal}}\frac{u_{\rm thermal}}{u_{\rm sim}}< 1.0$, 
which is the case for $\kappa>3$.

In the case of mixed $\kappa$ models we can calculate the particle acceleration efficiency as follows

\begin{equation}
\eta = \eta_{\rm acc}\frac{\eta_{\rm mod}}{1 + \eta_{\rm mod}} \frac{u_{\kappa}}{u_{\rm th}} \frac{u_{\rm th}}{u_{\rm sim}}
\end{equation}

\end{document}